\documentclass[oneeqnum,onefignum,onetabnum]{siamart251216}
\usepackage{amsmath,amssymb,bm}
\usepackage{graphicx,booktabs,algorithm,algorithmic,tikz}
\usepackage[section]{placeins}
\usepackage{flafter,needspace}
\usetikzlibrary{arrows.meta,positioning}
\hypersetup{colorlinks=true,allcolors=black,
pdfauthor={Aditya Dendukuri and Linda Petzold}}


\newcommand{\one}{\bm{1}}
\newcommand{\nnz}{\operatorname{nnz}}
\newcommand{\Cycle}{\operatorname{Cycle}}
\newsiamremark{example}{Example}
\headers{Multigrid for the RDME}{A. Dendukuri and L. Petzold}
\title{A Multigrid Finite State Projection Algorithm for the Solution
of the Reaction-Diffusion Master Equation}
\author{Aditya Dendukuri\thanks{Department of Computer Science,
University of California, Santa Barbara, USA.}
\and Linda Petzold\footnotemark[1]}
\begin{document}
\maketitle
\begin{abstract}
We develop a multigrid method for the finite state projection of the
reaction-diffusion master equation (RDME). The RDME evolves the joint
probabilities of molecular count configurations across well-mixed voxels
coupled by local reactions and diffusive jumps. When fast diffusion or
large reaction rates make this evolution stiff over long time horizons,
implicit time stepping requires solving a large shifted linear system at
every step. Grouping neighboring voxels and summing their molecular
counts produces a hierarchy of coarser state spaces terminating at a
one-voxel chemical master equation. Restriction sums residual entries
with the same coarse counts, while multinomial prolongation preserves the
RDME generator form on every level with explicit coarse reaction and
diffusion rates. For reversible diffusion, the error omitted by
coarsening is bounded by the graph Dirichlet energy of jumps inside the
aggregates, with a constant independent of population. Numerical tests
with standard multigrid cycles and point or block Gauss--Seidel smoothing
reach 7.16 million states and resolve rare probabilities of order
$10^{-7}$.
\end{abstract}

\section{Introduction}\label{sec:intro}
The reaction-diffusion master equation (RDME) models stochastic chemical
reactions and diffusion through the probabilities of discrete molecular
counts~\cite{engblom2009,isaacson2009}. The spatial domain is partitioned
into well-mixed cells, called voxels, and a state specifies the integer
copy number of each chemical species in each voxel. Molecules react
locally within voxels and jump between neighboring voxels by diffusion.
Because the master equation assigns a separate probability unknown to
every reachable arrangement of molecules across the grid, even a small
spatial domain can produce a large state space. Finite state projection
(FSP) retains a finite subset of configurations and treats transitions
leaving that subset as absorbing losses~\cite{munsky2006}, yielding a
finite linear evolution equation $\dot{\boldsymbol p}=\mathbf A\boldsymbol p$.

When fast diffusive jumps or large reaction propensities make this
evolution stiff over long time horizons, advancing the FSP probability
vector with an implicit time integrator requires solving a shifted linear
system $(\sigma I-\mathbf A)x=b$ at every step. Although the generator
$\mathbf A$ is sparse, the transition graph on molecular count
configurations has much higher connectivity than a physical mesh, and
sparse direct factorization can require far more memory than the matrix
and solution vector themselves. Multigrid methods avoid this
factorization bottleneck by combining local relaxation with corrections
computed on a hierarchy of smaller linear
systems~\cite{brandt1977,trottenberg2001}.

\subsection{Past Work}\label{sec:past-work}
Several numerical representations reduce the cost of integrating the
chemical master equation, including adaptive discrete Galerkin
approximations~\cite{deuflhard2008}, sparse
grids~\cite{hegland_sparse2008}, quantized tensor train
decompositions~\cite{kazeev2014}, and state aggregation~\cite{hegland2008}.
Multigrid and multilevel aggregation methods for Markov chains have
primarily targeted stationary distributions through multiplicative or
additive coarse corrections on the transition
graph~\cite{desterck2010,treister2011}. Our solver instead addresses the
shifted linear systems arising in implicit FSP time integration, pairing
voxel aggregation with standard V-, W-, and K-cycles~\cite{notay2008,trottenberg2001}
and point or block Gauss--Seidel smoothing~\cite{saad2003}.

\subsection{Contributions}\label{sec:contributions}
We construct a multigrid method for the finite RDME equations using
groups of physical voxels to determine the coarse states. Because each
fine unknown belongs to a complete spatial arrangement of molecules rather
than to a single spatial node, spatial coarsening must also specify how
probability entries are restricted and prolonged across compatible
arrangements. The contributions of this paper are as follows.
\begin{itemize}
\item We construct successive coarse state spaces by grouping neighboring
voxels and summing their molecular counts, terminating at a chemical
master equation (CME) on one voxel. Restriction sums residual entries
sharing the same coarse counts, and prolongation distributes coarse
corrections among compatible fine arrangements using multinomial weights.
These transfers supply additive corrections to the fine probability
vector, which remains the numerical solution throughout the iteration.
\item We derive closed-form coarse reaction and diffusion propensities by
averaging fine transitions under the multinomial weights. The Galerkin
coarse operator retains the reaction-diffusion generator structure and
absorbing boundary losses, allowing the same construction to be repeated
at every level without explicit sparse matrix multiplication.
\item For reversible diffusion, we bound the error omitted by the coarse
representation in terms of a weighted Dirichlet energy on the graph of
molecular arrangements. The resulting approximation constant is
independent of the molecular population for fixed aggregates and jump
rates, whereas a convergence bound for the full cycle would also require
estimates for smoothing and recursive coarse solves.
\item We examine convergence under population and spatial refinement and
compare the computational cost of standard cycles with point and block
Gauss--Seidel smoothing. Transient integrations reach 7.16 million
retained states, and a calculation of rare switching between metastable
states resolves an endpoint probability of order $10^{-7}$ that remains
poorly sampled across 60 million independent trajectories of the
stochastic simulation algorithm (SSA).
\end{itemize}
The coarse counts do not generally evolve as an autonomous Markov
chain~\cite{kemeny1976}, because nonlinear reaction propensities depend
on how molecules are distributed among the constituent fine voxels. The
coarse equations therefore serve strictly to compute corrections for the
fine linear system rather than to replace the fine evolution.

Section~\ref{sec:background} introduces the state probabilities, finite
state projection, and implicit time discretization.
Section~\ref{sec:method} presents the multigrid hierarchy, transfer
operators, smoothers, and time-integration algorithms.
Section~\ref{sec:analysis} states the Galerkin rate theorem and the
approximation and residual bounds, with complete proofs given in
Appendix~\ref{app:proofs}. The numerical experiments in
Section~\ref{sec:experiments} address the specified RDME rather than its
continuum limit as voxel volume vanishes~\cite{isaacson2009,hellander2012}.
The supplementary material contains a worked two-voxel calculation,
additional multilevel analysis, validation checks, implementation
measurements, and an SSA study of a propagating spatial front.

\section{States and Probability Equations}\label{sec:background}
We begin with molecular counts in a well-mixed reaction compartment and
then introduce a separate count for each species in each voxel. Finite
state projection selects the states whose probabilities are evolved, and
implicit time discretization produces the shifted linear systems solved
by multigrid.

\subsection{The RDME on One Voxel}\label{sec:cme}
On one well-mixed voxel of volume $\omega>0$, the RDME reduces to the
chemical master equation (CME). Consider ${N_{\mathrm s}}$ chemical
species $\mathrm{S}_1,\ldots,\mathrm{S}_{N_{\mathrm s}}$ in a well-mixed
volume $\omega>0$. A state records their molecular counts,
\[
 X=(X_1,\ldots,X_{N_{\mathrm s}})\in\mathbb N_0^{N_{\mathrm s}}.
\]
Let $p(X,t)$ be the probability of state $X$ at time $t$ and collect
these probabilities into the vector $\boldsymbol p(t)$. A state $X$ has
${N_{\mathrm s}}$ integer coordinates, whereas the probability vector
$\boldsymbol p(t)$ has one entry for every admissible combination of
counts, including states that carry zero probability initially. Each
chemical reaction alters the molecular counts and induces a linear
probability flux between the corresponding entries of $\boldsymbol p(t)$.

In reaction channel $r$, let $\alpha_{r,s}$ and $\beta_{r,s}$ denote the
numbers of molecules of species $s$ consumed and produced, respectively,
and let $c_r\ge0$ be the rate constant. Writing $\alpha_r=(\alpha_{r,s})$
and $\beta_r=(\beta_{r,s})$, the reaction channel and its mass-action
propensity are
\begin{equation}\label{eq:reaction-system}
 \sum_s\alpha_{r,s}\mathrm{S}_s\xrightarrow{c_r}\sum_s\beta_{r,s}\mathrm{S}_s,
 \qquad
 a_r(X)=c_r\omega^{1-q_r}\prod_s\binom{X_s}{\alpha_{r,s}},
 \qquad q_r=\sum_s\alpha_{r,s}.
\end{equation}
The propensity $a_r(X)$ is the transition rate from state $X$ to state
$X+\beta_r-\alpha_r$. Here $q_r$ is the reaction order, and the binomial
factors count the distinct combinations of reactant molecules available
in state $X$~\cite{gillespie1992}, with combinatorial rate factors
absorbed into $c_r$.

The probability flux through reaction $r$ out of state $X$ is the product
$a_r(X)p(X,t)$. Summing incoming fluxes from predecessor states
$X-\beta_r+\alpha_r$ and subtracting outgoing fluxes gives the master
equation
\begin{equation}\label{eq:cme-balance}
 \frac{d}{dt}p(X,t)
 =\sum_r\left[
 a_r(X-\beta_r+\alpha_r)p(X-\beta_r+\alpha_r,t)
 -a_r(X)p(X,t)\right],
\end{equation}
where terms with negative molecular counts are omitted. Although the
propensities $a_r(X)$ depend nonlinearly on the counts when $q_r>1$, the
balance \eqref{eq:cme-balance} is linear in the probabilities.
Enumerating the states as $X^{(i)}$ and letting $p_i(t)=p(X^{(i)},t)$
gives the matrix equation
\begin{equation}\label{def:cme}
 \dot{\boldsymbol p}=\mathbf A\boldsymbol p,
 \qquad\one^\top\mathbf A=0,
\end{equation}
where $\one$ is the vector of ones and $\mathbf A$ is the Markov
generator. Column $i$ of $\mathbf A$ collects the outgoing transitions
from source state $X^{(i)}$, with entries
\[
 \mathbf A_{ji}=\sum_{r:\,X^{(j)}=X^{(i)}+\beta_r-\alpha_r}a_r(X^{(i)})
 \quad(j\ne i),\qquad
 \mathbf A_{ii}=-\sum_{j\ne i}\mathbf A_{ji}.
\]
The off-diagonal entries are nonnegative transition rates, and each
diagonal entry equals the negative sum of the off-diagonal entries in its
column so that total probability is conserved.

\subsection{The RDME on Multiple Voxels}\label{sec:rdme}
The spatial RDME retains the linear probability balance of
\eqref{eq:cme-balance} while assigning each species a separate count in
every voxel. Consider an undirected graph of ${N_{\mathrm v}}$ well-mixed
voxels with positive volumes $\omega_v$. Since each voxel contains the
same ${N_{\mathrm s}}$ species, a spatial state is the count array
\begin{equation}\label{eq:spatial-state}
 X=(X_{v,s})=
 \begin{pmatrix}
 X_{1,1}&\cdots&X_{1,{N_{\mathrm s}}}\\
 \vdots&&\vdots\\
 X_{{N_{\mathrm v}},1}&\cdots&X_{{N_{\mathrm v}},{N_{\mathrm s}}}
 \end{pmatrix}\in\mathbb N_0^{{N_{\mathrm v}}\times {N_{\mathrm s}}}.
\end{equation}
The probability $p(X,t)$ belongs to this complete array, whose rows
specify the molecular counts in individual voxels. Summing $p(X,t)$ over
all rows except row $v$ yields the marginal distribution in voxel $v$,
but these voxel marginals do not form a closed system when nonlinear
reactions are present. The RDME evolves the joint distribution over all
reachable arrays \eqref{eq:spatial-state}.

Local reactions act on a single row of \eqref{eq:spatial-state}. Reaction
$r$ in voxel $v$ changes $X_{v,s}$ to $X_{v,s}+\beta_{r,s}-\alpha_{r,s}$
for each species while leaving all other rows unchanged, with propensity
\begin{equation}\label{eq:voxel-reaction}
 a_{r,v}(X)=c_r\omega_v^{1-q_r}
                     \prod_s\binom{X_{v,s}}{\alpha_{r,s}}.
\end{equation}
Diffusive jumps couple two adjacent rows by transferring one molecule of
species $s$ from voxel $v$ to a neighboring voxel $w$. Each directed edge
$(v,w)$ and species $s$ gives the first-order transition
$\mathrm{S}_{v,s}\to \mathrm{S}_{w,s}$ acting on $(X_{v,s},X_{w,s})$ by
\[
 (X_{v,s},X_{w,s})\longrightarrow(X_{v,s}-1,X_{w,s}+1)
 \quad\text{at rate }d_{vw}^{(s)}X_{v,s}.
\]
The reverse jump from $w$ to $v$ occurs at rate $d_{wv}^{(s)}$ per
molecule, which may differ from $d_{vw}^{(s)}$ when voxel volumes are
unequal. Summing the reaction and diffusion fluxes according to
\eqref{eq:cme-balance} gives the spatial master equation
$\dot{\boldsymbol p}=\mathbf A\boldsymbol p$. Each configuration has
${N_{\mathrm v}}{N_{\mathrm s}}$ integer coordinates, and every elementary
reaction or diffusion event connects two configurations. With a single
voxel there are no diffusive jumps, and the RDME reduces to the CME.

\subsection{Finite State Projection}\label{sec:fsp}
When open reactions allow molecular counts to grow without bound, we
select a finite state set $\mathcal X$ and evolve its probability entries
by the finite state projection~\cite{munsky2006}. In our multigrid
construction, retention is specified by the species totals
$k_s=\sum_vX_{v,s}$. All spatial arrangements sharing the same totals
$k=(k_s)$ form a population block, and $\mathcal X$ is chosen as a finite
union of complete population blocks. Because diffusive jumps conserve
every species total $k_s$, no diffusion transition can leave
$\mathcal X$. Only reaction channels that change the species totals can
cross the truncation boundary. For one species truncated at maximum
population ${N_{\max}}$, the retained state space is
\begin{equation}\label{eq:retained-states}
 \mathcal X=\left\{X\in\mathbb N_0^{N_{\mathrm v}}:\sum_vX_v\le {N_{\max}}\right\},
 \qquad N=|\mathcal X|=\binom{{N_{\mathrm v}}+{N_{\max}}}{{N_{\max}}}.
\end{equation}

Bounding the total population $\sum_v X_v$ rather than capping each voxel
count $X_v$ separately makes $\mathcal X$ closed under diffusion.
Restricting the generator to the principal submatrix on these $N$ states
gives the finite evolution equation
\begin{equation}\label{eq:fsp}
 \dot{\boldsymbol p}=\mathbf A\boldsymbol p,\qquad
 \one^\top\mathbf A\le0,
\end{equation}
where $\mathbf A$ and $\boldsymbol p$ henceforth denote the retained
generator and probability vector. Omitting a destination state outside
$\mathcal X$ removes its off-diagonal entry while leaving its outgoing
rate in the source diagonal entry $\mathbf A_{ii}$, so escaping
probability mass is absorbed rather than reflected to a boundary state.

For a nonexplosive Markov chain whose initial distribution is supported in
$\mathcal X$, the exact solution of \eqref{eq:fsp} extended by zero
outside $\mathcal X$ has $\ell_1$ truncation error
$1-\one^\top\boldsymbol p(t)$~\cite{munsky2006}. This missing mass
measures truncation of the continuous-time chain independently of the
additional errors from time stepping and iterative linear solves. When a
fixed cap is insufficient over long integrations, adaptive FSP enlarges
$\mathcal X$ by appending complete population blocks and retries a
rejected step from the previous distribution, as detailed in
Supplementary Section~S3.2.

\subsection{Time Discretization}\label{sec:time}
We integrate the finite evolution equation \eqref{eq:fsp} by solving a
shifted linear system at each time step. Let
$\boldsymbol p^j\approx\boldsymbol p(j\Delta t)$ for step size
$\Delta t>0$. Both backward Euler and Crank--Nicolson advance
$\boldsymbol p^{j-1}$ to $x=\boldsymbol p^j$ by solving
\begin{equation}\label{eq:system}
 \mathbf A_\sigma x=b,\qquad\mathbf A_\sigma =\sigma I-\mathbf A,\qquad\sigma>0,
\end{equation}
where the shift $\sigma$ and right-hand side $b$ are given by
\begin{equation}\label{eq:time-steps}
 \begin{array}{c|cc}
 &\sigma&b\\ \hline
 \text{backward Euler}&1/\Delta t&\boldsymbol p^{j-1}/\Delta t\\[2pt]
 \text{Crank--Nicolson}&2/\Delta t&
 (2/\Delta t)\boldsymbol p^{j-1}+\mathbf A\boldsymbol p^{j-1}.
 \end{array}
\end{equation}
Here lowercase $x$ denotes a vector on $\mathcal X$, while uppercase $X$
denotes a molecular count configuration. Because $\mathbf A$ has
nonnegative off-diagonal entries and nonpositive column sums,
$\mathbf A_\sigma$ is a nonsingular $M$-matrix with column sums bounded
below by $\sigma>0$. Each iterative solve starts from the preceding
distribution $\boldsymbol p^{j-1}$ and applies multigrid cycles until the
residual meets the stopping tolerance. For Crank--Nicolson integrations,
we replace the first interval by two backward Euler half-steps of size
$\Delta t/2$. Both half-steps and the subsequent Crank--Nicolson steps
share the shift $\sigma=2/\Delta t$, allowing the assembled level
matrices and relaxation factors to be reused across steps.

\FloatBarrier
\section{The Multigrid Method}\label{sec:method}\label{sec:hierarchy}
At each time step, multigrid solves \eqref{eq:system} by adding
corrections to an approximate fine probability vector $x$. The discrepancy
of $x$ in \eqref{eq:system} is the residual $r$, and the unknown algebraic
error $e$ satisfies the residual equation
\[
 r=b-\mathbf A_\sigma x,\qquad e=\mathbf A_\sigma^{-1}b-x,\qquad\mathbf A_\sigma e=r.
\]
Although $e$ is unknown, the residual $r$ is directly computable from
$x$, $b$, and $\mathbf A_\sigma$. Multigrid approximates $e$ by combining
local relaxation on the fine space with corrections computed on a
hierarchy of coarser state spaces.

To motivate the choice of coarse spaces, consider the diffusion generator
$\mathbf A_{\mathrm d}$ inside $\mathbf A$, whose transitions connect
molecular arrangements sharing the same species totals. When the diffusive
jump rates are reversible with respect to voxel volumes,
$\mathbf A_{\mathrm d}$ admits positive equilibrium weights $\pi_i$ on
each population block satisfying
$\pi_i(\mathbf A_{\mathrm d})_{ji}=\pi_j(\mathbf A_{\mathrm d})_{ij}$.
Section~\ref{sec:approximation} constructs these weights explicitly from
the voxel volumes without assuming equilibrium for the reacting system.
Because the equilibrium weights $\pi_i$ vary across arrangements, we
measure smoothness of an error vector $e$ through its relative entries
$f_i=e_i/\pi_i$. The associated graph Dirichlet energy and weighted norm
are
\begin{equation}\label{eq:diffusion-energy}
 \frac12\sum_{i\ne j}\pi_i(\mathbf A_{\mathrm d})_{ji}(f_i-f_j)^2,
 \qquad \|e\|_{\pi^{-1}}^2=\sum_i\frac{e_i^2}{\pi_i}.
\end{equation}
This graph energy is small when the relative error $f$ varies slowly
across diffusive jumps. For an eigenmode of $-\mathbf A_{\mathrm d}$, its
graph energy divided by $\|e\|_{\pi^{-1}}^2$ equals its decay rate under
pure diffusion, so modes with a small energy ratio decay slowly and form
the analogue of smooth geometric modes for the probability equation.

Our coarse spaces group neighboring voxels into aggregates and retain
their combined molecular counts, representing errors whose relative
entries $f_i$ vary little among arrangements that share the same
aggregate counts. Relaxation supplies fine updates within each aggregate
class that cannot be represented through coarse counts alone.
Section~\ref{sec:approximation} bounds the approximation error omitted by
coarsening in terms of the portion of \eqref{eq:diffusion-energy} coming
from diffusive jumps inside the aggregates.

Repeating the voxel aggregation recursively reduces the grid until a
single voxel remains. A multigrid cycle descends through the hierarchy by
applying forward relaxation followed by residual restriction. At the
bottom level, the shifted CME on one voxel is solved directly to obtain a
coarse correction. The cycle then ascends through the hierarchy,
prolonging and adding each correction before applying reverse relaxation
at every level. Throughout the cycle, the fine probability vector remains
the primary numerical unknown, and the coarse equations serve solely to
accelerate the fine linear solve.

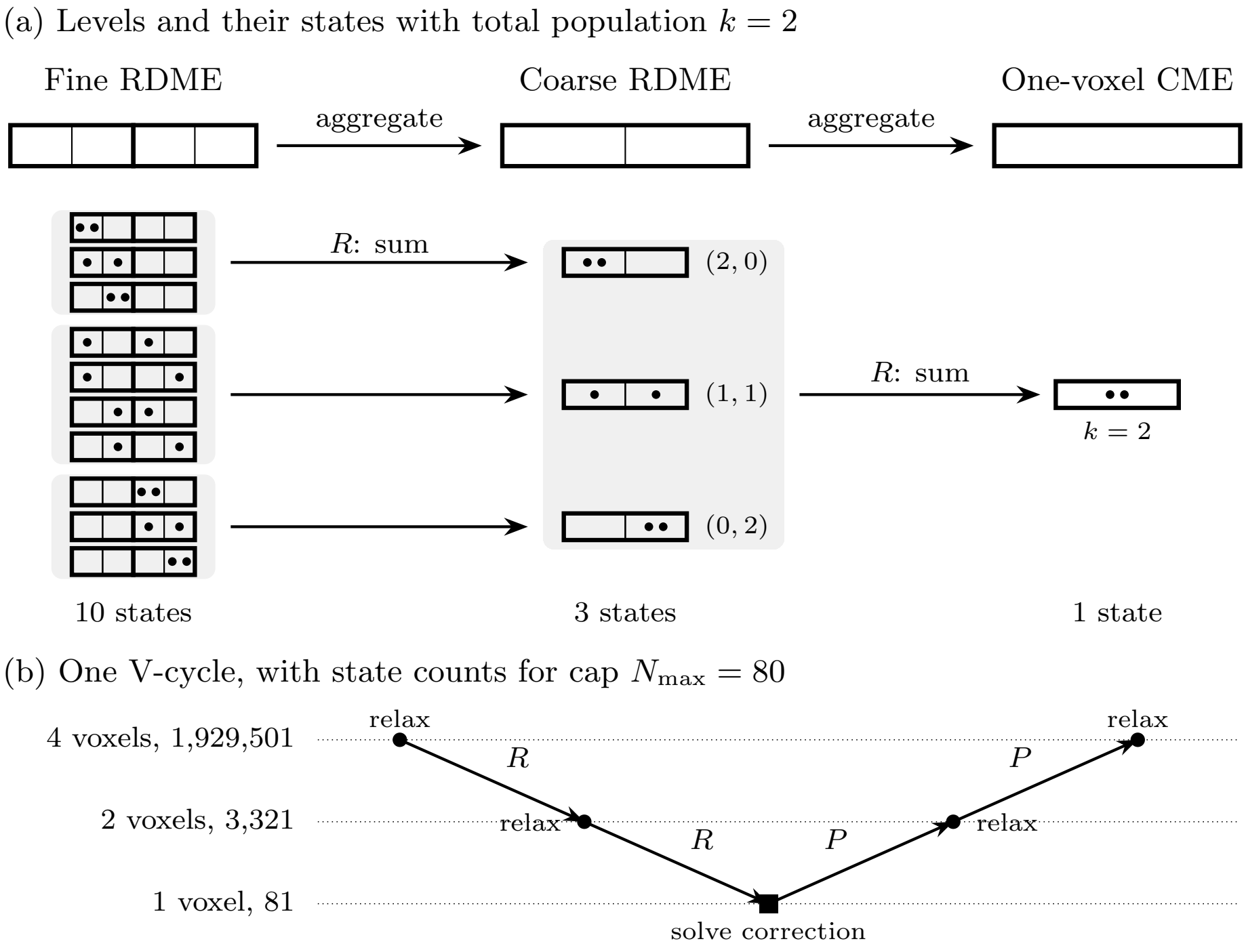
\begin{figure}[!htbp]
\centering
\begin{tikzpicture}[x=1cm,y=1cm,>=Stealth,font=\small]
% \strip{x0}{yc}{cell width}{counts}{cells per aggregate}{cells}: one arrangement,
% with dots for molecules and heavy outlines on the next aggregates.
\def\sh{.13}\def\dotr{.042}
\def\strip#1#2#3#4#5#6{%
 \edef\stripl{#4}\foreach \c [count=\i from 0] in \stripl{%
  \draw (#1+\i*#3,#2-\sh) rectangle ++(#3,2*\sh);
  \ifnum\c=1 \fill (#1+\i*#3+.5*#3,#2) circle (\dotr);\fi
  \ifnum\c=2 \fill (#1+\i*#3+.5*#3-.07,#2) circle (\dotr)
                   (#1+\i*#3+.5*#3+.07,#2) circle (\dotr);\fi}
 \pgfmathtruncatemacro\nag{#6/#5-1}%
 \foreach \j in {0,...,\nag}
  \draw[very thick] (#1+\j*#5*#3,#2-\sh) rectangle ++(#5*#3,2*\sh);}
% (a) Voxel grids and the states they distinguish.
\node[anchor=west] at (-.2,1.0) {(a) Levels and their states with total population $k=2$};
\node at (1.2,.45) {Fine RDME};
\node at (6.0,.45) {Coarse RDME};
\node at (10.8,.45) {One-voxel CME};
\draw (0,-.4) rectangle (2.4,0);
\foreach \x in {.6,1.2,1.8} \draw (\x,-.4)--(\x,0);
\draw[very thick] (0,-.4) rectangle (1.2,0);
\draw[very thick] (1.2,-.4) rectangle (2.4,0);
\draw (4.8,-.4) rectangle (7.2,0);
\draw (6,-.4)--(6,0);
\draw[very thick] (4.8,-.4) rectangle (7.2,0);
\draw[very thick] (9.6,-.4) rectangle (12,0);
\draw[->,thick] (2.6,-.2)--node[above,font=\footnotesize] {aggregate} (4.6,-.2);
\draw[->,thick] (7.4,-.2)--node[above,font=\footnotesize] {aggregate} (9.4,-.2);
% Fine states, grouped by their coarse counts.
\fill[gray!12,rounded corners=3pt] (.4,-.83) rectangle (2.0,-1.85);
\fill[gray!12,rounded corners=3pt] (.4,-1.95) rectangle (2.0,-3.31);
\fill[gray!12,rounded corners=3pt] (.4,-3.41) rectangle (2.0,-4.43);
\foreach \y/\n in {-1.0/{2,0,0,0},-1.34/{1,1,0,0},-1.68/{0,2,0,0},
  -2.12/{1,0,1,0},-2.46/{1,0,0,1},-2.80/{0,1,1,0},-3.14/{0,1,0,1},
  -3.58/{0,0,2,0},-3.92/{0,0,1,1},-4.26/{0,0,0,2}}
 \strip{.6}{\y}{.3}{\n}{2}{4}
% Coarse states, all with the same total.
\fill[gray!12,rounded corners=3pt] (5.2,-1.12) rectangle (7.55,-4.14);
\foreach \y/\n/\lab in {-1.34/{2,0}/{(2,0)},-2.63/{1,1}/{(1,1)},-3.92/{0,2}/{(0,2)}}{
 \strip{5.4}{\y}{.6}{\n}{2}{2}
 \node[font=\scriptsize,anchor=west] at (6.65,\y) {$\lab$};
 \draw[->,thick] (2.15,\y)--(5.05,\y);}
\node[font=\footnotesize] at (3.6,-1.15) {$R$: sum};
% One-voxel state.
\strip{10.2}{-2.63}{1.2}{2}{1}{1}
\node[font=\scriptsize] at (10.8,-2.98) {$k=2$};
\draw[->,thick] (7.7,-2.63)--node[above,font=\footnotesize] {$R$: sum} (10.05,-2.63);
\node[font=\footnotesize] at (1.2,-4.75) {10 states};
\node[font=\footnotesize] at (6.0,-4.75) {3 states};
\node[font=\footnotesize] at (10.8,-4.75) {1 state};
% (b) One V-cycle through the same levels.
\node[anchor=west] at (-.2,-5.35) {(b) One V-cycle, with state counts for cap ${N_{\max}}=80$};
\foreach \y/\lab in {-6.0/{4 voxels, 1{,}929{,}501},-6.8/{2 voxels, 3{,}321},-7.6/{1 voxel, 81}}{
 \draw[densely dotted,line width=.3pt] (3.0,\y)--(12,\y);
 \node[font=\footnotesize,anchor=east] at (2.9,\y) {\lab};}
\draw[-{Stealth[length=2mm]},thick] (3.8,-6.0)
 --node[above right,font=\footnotesize] {$R$} (5.6,-6.8);
\draw[-{Stealth[length=2mm]},thick] (5.6,-6.8)
 --node[above right,font=\footnotesize] {$R$} (7.4,-7.6);
\draw[-{Stealth[length=2mm]},thick] (7.4,-7.6)
 --node[above left,font=\footnotesize] {$P$} (9.2,-6.8);
\draw[-{Stealth[length=2mm]},thick] (9.2,-6.8)
 --node[above left,font=\footnotesize] {$P$} (11.0,-6.0);
\foreach \x/\y in {3.8/-6.0,5.6/-6.8,9.2/-6.8,11.0/-6.0}
 \fill (\x,\y) circle (2pt);
\fill (7.31,-7.69) rectangle (7.49,-7.51);
\node[font=\scriptsize,above] at (3.8,-6.0) {relax};
\node[font=\scriptsize,left] at (5.5,-6.8) {relax};
\node[font=\scriptsize,below] at (7.4,-7.65) {solve correction};
\node[font=\scriptsize,right] at (9.3,-6.8) {relax};
\node[font=\scriptsize,above] at (11.0,-6.0) {relax};
\end{tikzpicture}
\caption{Hierarchy for one species from a four-voxel RDME to a one-voxel CME.
(a) Heavy outlines group voxels for the next aggregation, and dots
are molecules. The 10 arrangements of two molecules map to 3 coarse
states and then the single CME state $k=2$. Restriction sums vector
entries over each shaded group. (b) A V-cycle relaxes and restricts
the residual on descent, solves the correction equation directly
at the bottom, and prolongs the correction with reverse relaxation
on ascent. State counts are those of the Schl\"ogl test in
Section~\ref{sec:cycle-results} at cap 80.}
\label{fig:hierarchy-overview}
\end{figure}

Sections~\ref{sec:transfer}--\ref{sec:rates} introduce the coarse
states, transfer operators, and Galerkin coarse equations.
Sections~\ref{sec:smoothing}--\ref{sec:algorithms} describe the smoothers,
cycle recursions, stopping rule, and complete time-integration algorithms.
Figure~\ref{fig:hierarchy-overview} illustrates the state hierarchy and a
single V-cycle, and Supplementary Section~S1
presents a worked two-voxel calculation.

\subsection{Coarse States and Restriction}\label{sec:transfer}
Restriction groups fine states by their molecular counts on aggregated
voxels and sums the corresponding vector entries. Consider a partition of
the voxel graph into disjoint connected groups $C_k$, called aggregates.
Each aggregate $C_k$ forms one coarse voxel with volume and species
counts
\begin{equation}\label{eq:coarse-counts}
 \Omega_k=\sum_{v\in C_k}\omega_v,\qquad
 {X^c}_{k,s}(X)=\sum_{v\in C_k}X_{v,s}.
\end{equation}
Let ${X^c}(X)$ denote the resulting coarse configuration and let
$\mathcal X_c=\{{X^c}(X):X\in\mathcal X\}$ denote the coarse state space,
with dimension $N_c=|\mathcal X_c|$. Enumerating the fine and coarse
states as $X^{(i)}$ and $X^{c,(j)}$, two fine states map to the same
coarse state whenever their aggregate counts agree, regardless of how the
molecules are arranged inside each aggregate. The restriction matrix
$R\in\mathbb R^{N_c\times N}$ acts by
\begin{equation}\label{eq:transfer-action}
 (Rx)_j=\sum_{i:\,X^c(X^{(i)})=X^{c,(j)}}x_i,\qquad
 R_{ji}=\begin{cases}1,&X^c(X^{(i)})=X^{c,(j)},\\0,&\text{otherwise}.\end{cases}
\end{equation}
Applied to a fine probability distribution $\boldsymbol p$, restriction
computes the exact coarse marginal distribution
$\boldsymbol p_c=R\boldsymbol p$. Within a multigrid cycle, $R$ acts on
the fine residual $r$, summing its signed entries over each coarse
equivalence class.

Summing counts over each aggregate preserves the global species totals
$k_s=\sum_v X_{v,s}$ while removing distinctions between internal voxel
arrangements. The construction permits any partition into connected
aggregates, and our Cartesian experiments group $2^d$ adjacent voxels in
$d$ dimensions.

\subsection{Prolongation to Fine Arrangements}\label{sec:prolongation}
To distribute a coarse correction among fine arrangements with prescribed
aggregate counts, we place the ${X^c}_{k,s}$ molecules of species $s$ in
aggregate $C_k$ independently across its constituent voxels, choosing
$v\in C_k$ with probability $\omega_v/\Omega_k$. These volume fractions
give uniform expected concentration across the voxels of each aggregate.
Multiplying the placement probabilities by the multinomial count of
indistinguishable permutations gives, for fine state $X=X^{(i)}$ and
coarse state $X^c=X^{c,(j)}$, the prolongation weight
\begin{equation}\label{eq:P}
 P_{ij}=\prod_{k,s}\left[
   \frac{{X^c}_{k,s}!}{\prod_{v\in C_k}X_{v,s}!}
   \prod_{v\in C_k}\left(\frac{\omega_v}{\Omega_k}\right)^{X_{v,s}}
 \right]\quad\text{if }{X^c}(X)={X^c},
\end{equation}
and $P_{ij}=0$ otherwise. The weight factors across aggregates and
species because their placements are independent.
Section~\ref{sec:approximation} shows that $P_{ij}$ equals the
diffusion equilibrium probability conditioned on the aggregate counts
when the jump rates satisfy detailed balance with the voxel volumes. For
a coarse vector $x_c\in\mathbb R^{N_c}$, prolongation
$P\in\mathbb R^{N\times N_c}$ acts by
\begin{equation}\label{eq:prolongation-action}
 (Px_c)_i=\sum_jP_{ij}(x_c)_j.
\end{equation}
Each column of $R$ contains a single nonzero entry equal to one, and each
row of $P$ contains a single nonzero entry equal to its multinomial
weight $P_{ij}$.

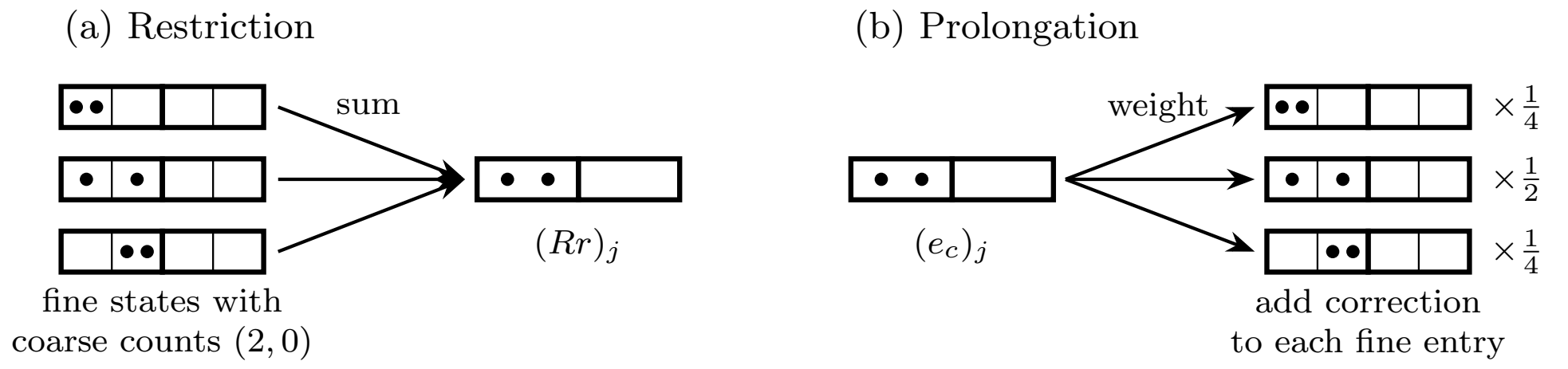
\begin{figure}[!htbp]
\centering
\begin{tikzpicture}[x=1cm,y=1cm,>=Stealth,font=\small]
% \strip{x0}{yc}{cell width}{counts}{cells per aggregate}{cells}: one arrangement,
% with dots for molecules and heavy outlines on aggregates.
\def\sh{.17}\def\dotr{.055}
\def\strip#1#2#3#4#5#6{%
 \edef\stripl{#4}\foreach \c [count=\i from 0] in \stripl{%
  \draw (#1+\i*#3,#2-\sh) rectangle ++(#3,2*\sh);
  \ifnum\c=1 \fill (#1+\i*#3+.5*#3,#2) circle (\dotr);\fi
  \ifnum\c=2 \fill (#1+\i*#3+.3*#3,#2) circle (\dotr)
                   (#1+\i*#3+.7*#3,#2) circle (\dotr);\fi}
 \pgfmathtruncatemacro\nag{#6/#5-1}%
 \foreach \j in {0,...,\nag}
  \draw[very thick] (#1+\j*#5*#3,#2-\sh) rectangle ++(#5*#3,2*\sh);}
% (a) Restriction
\node[anchor=west] at (-.1,1.25) {(a) Restriction};
\strip{0}{.6}{.42}{2,0,0,0}{2}{4}
\strip{0}{0}{.42}{1,1,0,0}{2}{4}
\strip{0}{-.6}{.42}{0,2,0,0}{2}{4}
\foreach \y in {.6,0,-.6}
 \draw[->,thick] (1.8,\y)--(3.35,0);
\node[font=\footnotesize] at (2.55,.6) {sum};
\strip{3.45}{0}{.84}{2,0}{1}{2}
\node[font=\footnotesize] at (4.29,-.55) {$(Rr)_j$};
\node[font=\footnotesize,align=center] at (.84,-1.2)
 {fine states with\\coarse counts $(2,0)$};
% (b) Prolongation
\begin{scope}[xshift=6.55cm]
\node[anchor=west] at (-.1,1.25) {(b) Prolongation};
\strip{0}{0}{.84}{2,0}{1}{2}
\node[font=\footnotesize] at (.84,-.55) {$(e_c)_j$};
\foreach \y/\w in {.6/\tfrac14,0/\tfrac12,-.6/\tfrac14}{
 \draw[->,thick] (1.78,0)--(3.35,\y);
 \node[font=\footnotesize,anchor=west] at (5.18,\y) {$\times\w$};}
\node[font=\footnotesize] at (2.55,.6) {weight};
\strip{3.45}{.6}{.42}{2,0,0,0}{2}{4}
\strip{3.45}{0}{.42}{1,1,0,0}{2}{4}
\strip{3.45}{-.6}{.42}{0,2,0,0}{2}{4}
\node[font=\footnotesize,align=center] at (4.29,-1.2)
 {add correction\\to each fine entry};
\end{scope}
\end{tikzpicture}
\caption{The transfer operations for one species with coarse state $X^c=(2,0)$
at index $j$ on four voxels of equal volume with aggregates $\{v_1,v_2\}$ and
$\{v_3,v_4\}$. Dots are molecules and heavy outlines are aggregates.
(a) Restriction sums the residual entries of the three fine
arrangements with these counts. (b) Prolongation distributes the
coarse correction by the multinomial weights \eqref{eq:P} and adds
it to the fine iterate.}
\label{fig:transfer-operations}
\end{figure}

Because the transfer operators preserve species totals, ordering states
by total population and then grouping fine states by their coarse image
$X^{c,(j)}$ puts both matrices into block-diagonal form. Letting $w_j$
denote the column vector of multinomial weights \eqref{eq:P} for fine
states in class $j$, this ordering gives
\begin{equation}\label{eq:transfer-blocks}
 R=\begin{pmatrix}
 \one^\top&0&\cdots\\
 0&\one^\top&\cdots\\
 \vdots&\vdots&\ddots
 \end{pmatrix},\qquad
 P=\begin{pmatrix}
 w_1&0&\cdots\\
 0&w_2&\cdots\\
 \vdots&\vdots&\ddots
 \end{pmatrix}.
\end{equation}
In each block $j$, the row vector $\one^\top$ sums the entries of the
compatible fine arrangements, and the column vector $w_j$ distributes the
coarse correction across them with $\one^\top w_j=1$, as illustrated in
Figure~\ref{fig:transfer-operations}.

Because each fine state belongs to a unique coarse class and the weights
in each column of $P$ sum to unity by the multinomial theorem, the
transfer operators satisfy
\begin{equation}\label{eq:transfer-identities}
 RP=I,\qquad\one^\top R=\one^\top,\qquad
 \one^\top P=\one^\top,
\end{equation}
where each vector $\one$ has compatible dimension. Restriction therefore
recovers any prolonged coarse vector, and $(PR)^2=P(RP)R=PR$ makes $PR$ a
projection on the fine space.

\subsection{Coarse Correction}\label{sec:systems}
A coarse-level correction has the prolonged form $Pe_c$, and its action on
the restricted residual is $R\mathbf A_\sigma P e_c$. Requiring this
term to cancel the restricted residual $Rr$ determines the Galerkin coarse
system and fine update
\begin{equation}\label{eq:correction}
 \mathbf A_{\sigma,c}e_c=Rr,\qquad x^+=x+Pe_c,
 \qquad\mathbf A_{\sigma,c}=R\mathbf A_\sigma P.
\end{equation}
The coarse correction $e_c$ has signed entries. An exact coarse solve
annihilates the restricted residual,
\[
 R(b-\mathbf A_\sigma x^+)=Rr-\mathbf A_{\sigma,c}e_c=0,
\]
so the updated iterate $x^+$ satisfies the fine equation after summation
over each coarse count class. Because $RP=I$ implies $(I-PR)P=0$, the
update also preserves the fine component omitted by restriction,
\begin{equation}\label{eq:fine-component}
 (I-PR)x^+=(I-PR)x,
\end{equation}
adding a coarse correction without replacing the fine iterate by a
multinomial distribution. Relaxation damps error components in the null
space of $R$ before and after \eqref{eq:correction}.

\subsection{Coarse Equations}\label{sec:rates}
To apply \eqref{eq:correction} recursively across multiple levels, we
assemble the Galerkin coarse generator
\begin{equation}\label{eq:coarse-generator}
 \mathbf A_c=R\mathbf AP.
\end{equation}
For distinct coarse states $j\ne k$, expanding the matrix product shows
that $(\mathbf A_c)_{jk}$ averages all fine transition rates leading from
class $k$ to class $j$,
\begin{equation}\label{eq:coarse-entry}
 (\mathbf A_c)_{jk}
 =\sum_{i:\,R_{ki}=1}P_{ik}
          \sum_{m:\,R_{jm}=1}\mathbf A_{mi},\qquad j\ne k.
\end{equation}
Evaluating these averages analytically yields closed-form coarse
diffusion and reaction propensities without forming the sparse matrix
product $R\mathbf AP$.

For diffusion (suppressing the species index), a molecule in aggregate
$C_k$ occupies fine voxel $v\in C_k$ with marginal probability
$\omega_v/\Omega_k$ under \eqref{eq:P}, so its expected jump rate per
molecule into another aggregate $C_j$ ($j\ne k$) is
\begin{equation}\label{eq:coarse-diffusion}
 d^c_{kj}=\sum_{v\in C_k}\frac{\omega_v}{\Omega_k}
                         \sum_{w\in C_j}d_{vw},\qquad k\ne j,
\end{equation}
with coarse diffusive propensity $d^c_{kj}{X^c}_k$. Diffusive jumps
between fine voxels inside the same aggregate leave the coarse counts
unchanged and cancel between off-diagonal and diagonal entries. On a
uniform $d$-dimensional Cartesian grid of spacing $h$ and jump rate
$D/h^2$ coarsened by grouping $2^d$ voxels per aggregate, adjacent
aggregates share $2^{d-1}$ fine interface edges with
$\omega_v/\Omega_k=2^{-d}$, so \eqref{eq:coarse-diffusion} gives
$d^c_{kj}=2D/(2h)^2$, which is twice the naive coarse finite-difference
rate $D/(2h)^2$.

If the reaction rate constant $c_r$ is uniform across the fine voxels of
aggregate $C_k$, the corresponding coarse reaction propensity is
\begin{equation}\label{eq:coarse-reaction}
 a^c_{r,k}({X^c})=c_r\Omega_k^{1-q_r}
                         \prod_s\binom{{X^c}_{k,s}}{\alpha_{r,s}}.
\end{equation}
Theorem~\ref{thm:reaction} proves that \eqref{eq:coarse-diffusion} and
\eqref{eq:coarse-reaction}, together with absorbing losses at omitted
destination states, reproduce the Galerkin generator $\mathbf A_c=R\mathbf AP$.

Because $\mathbf A_c$ retains the algebraic form of an RDME generator on
the aggregate voxel graph, repeating the construction across
$N_{\mathrm l}$ coarsening steps gives the level generators
\begin{equation}\label{eq:levels}
 \mathbf A_0=\mathbf A,\qquad
 \mathbf A_{ l +1}=R_l\mathbf A_l P_l,
 \qquad l =0,\ldots,{N_{\mathrm l}}-1,
\end{equation}
provided each reaction rate constant $c_r$ is uniform within every
aggregate at every level. The bottom level ${N_{\mathrm l}}$ has a single
voxel of volume $\Omega=\sum_v\omega_v$ and no diffusion, reducing to a
well-mixed CME on the species totals with ${N_{\max}}+1$ unknowns for one
species at cap ${N_{\max}}$. By \eqref{eq:transfer-identities}, the
shifted coarse operators satisfy
\begin{equation}\label{eq:shifted-coarse}
 \mathbf A_{\sigma,c}=R(\sigma I-\mathbf A)P=\sigma I-\mathbf A_c,
\end{equation}
so the same generator hierarchy $\{\mathbf A_l\}_{l=0}^{N_{\mathrm l}}$
applies at every positive shift $\sigma$.

\subsection{Smoothing}\label{sec:relaxation}\label{sec:smoothing}
Because a coarse correction $Pe_c$ lies in the column space of $P$, it
cannot independently adjust relative errors among fine arrangements that
share the same coarse counts. Relaxation sweeps on $\mathbf A_\sigma x=b$
supply these local updates before and after coarse correction. We apply
point or approximate block Gauss--Seidel to the complete shifted matrix
$\mathbf A_\sigma$, including reactions, diffusion, absorbing losses, and
the shift. Point Gauss--Seidel updates one state component at a time via
\begin{equation}\label{eq:point-sweep}
 x_i\leftarrow x_i+
      \frac{b_i-(\mathbf A_\sigma x)_i}{(\mathbf A_\sigma)_{ii}},
\end{equation}
using each updated entry immediately in subsequent rows. The default
point smoother applies two forward sweeps before coarse correction and
two reverse sweeps afterward.

Updating an entire population block simultaneously accounts for the
diffusive coupling among all arrangements of that total population. For a
single species with birth and death reactions that change the molecule
count by one, diffusion preserves the total population $k=\sum_v X_v$
while reactions change $k$ by $\pm1$, so ordering states by
$k=0,\ldots,{N_{\max}}$ makes $\mathbf A_\sigma$ block tridiagonal.
Letting $x_k$ and $b_k$ denote the subvectors for population $k$,
approximate block Gauss--Seidel updates each population block in turn via
\begin{equation}\label{eq:block-sweep}
 r_k=b_k-\sum_{j=k-1}^{k+1}(\mathbf A_\sigma)_{kj}x_j,
 \qquad x_k\leftarrow x_k+\widetilde{\mathbf A}_{\sigma,k}^{-1}r_k,
\end{equation}
where $\widetilde{\mathbf A}_{\sigma,k}$ is the zero-fill incomplete LU
factorization, denoted ILU(0), of the diagonal block
$(\mathbf A_\sigma)_{kk}$~\cite{saad2003} and indices outside
$0,\ldots,{N_{\max}}$ are omitted. The default block smoother applies one
forward sweep before coarse correction and one reverse sweep afterward.
On the bottom CME level $l={N_{\mathrm l}}$, each population block is a
scalar entry, and the resulting tridiagonal matrix is factored once in
linear time. For multi-species models, point relaxation and a sparse
direct bottom solver can be used without changing the transfer operators.

Because smoothing is separate from restriction, coarse solution, and
prolongation, replacing the smoother leaves $R_l$, $P_l$, and
$\mathbf A_l$ unchanged. For example, cascading smoothers~\cite{saye2026}
could be constructed for these shifted systems and applied at the same
points in the cycle.

\subsection{Multigrid Cycles}\label{sec:cycles}\label{sec:vw}
A multigrid cycle combines relaxation on level $l$ with a recursive
coarse correction on level $l+1$. Let $\Cycle( l ,b)$ denote one cycle
for $\mathbf A_{\sigma,l}x=b$ starting from a zero initial guess, where
$\mathbf A_{\sigma,l}=\sigma I-\mathbf A_l$. After forward relaxation of
$x$ on level $l$ and restriction of the residual
$b_c=R_l(b-\mathbf A_{\sigma,l} x)$, the V- and W-cycles compute $e_c$ by
\begin{equation}\label{eq:vw}
 e_c^{(0)}=0,\qquad
 e_c^{(j+1)}=e_c^{(j)}+
 \Cycle( l +1,b_c-\mathbf A_{\sigma,l +1}e_c^{(j)}),
\end{equation}
performing one recursive step ($e_c=e_c^{(1)}$) in a V-cycle and two
recursive steps ($e_c=e_c^{(2)}$) in a W-cycle~\cite{trottenberg2001}.

\phantomsection\label{sec:kcycle}
The K-cycle instead minimizes the coarse residual norm
$\|b_c-\mathbf A_{\sigma,l+1}e_c\|_2$ over at most two recursive
directions~\cite{notay2008}. It first scales the initial recursive
direction $v_1=\Cycle(l+1,b_c)$ to minimize the coarse residual norm
along $v_1$. If this scaled correction reduces the coarse residual norm
by at least a factor of four, it is accepted without a second call.
Otherwise, the cycle computes a second recursive direction on the
remaining residual and chooses the optimal linear combination of both
directions, as detailed in Supplementary Section~S2.1.
At the bottom level $l={N_{\mathrm l}}$, the direct CME solve replaces
this procedure. Because the optimal weights depend on the residual, the
K-cycle is nonlinear and is applied directly as a standalone solver,
whereas the linear W-cycle is used when preconditioning
GMRES~\cite{saad1986}.

\subsection{Accuracy and Computational Cost}\label{sec:stopping}\label{sec:storage}
At each time step, we initialize $x$ from the preceding probability
distribution and apply
\[
 x\leftarrow x+\Cycle(0,b-\mathbf A_{\sigma,0}x)
\]
until $\|b-\mathbf A_{\sigma,0}x\|_1/\sigma\le\varepsilon$.
Proposition~\ref{prop:residual} shows that this stopping rule bounds the
$\ell_1$ algebraic error $\|x-\mathbf A_{\sigma,0}^{-1}b\|_1$ by
$\varepsilon$ using only the computable fine residual. Time discretization
and FSP truncation errors are assessed separately.

Applying $R_l$ or $P_l$ requires $O(N_l)$ work, and a point sweep or an
application of stored ILU(0) factors requires $O(\nnz(\mathbf A_{\sigma,l}))$
work. For one species on a $d$-dimensional Cartesian hierarchy at fixed
cap ${N_{\max}}\ge1$, summing this work across levels yields $O(N)$ cost
per V-, W-, or K-cycle when $d{N_{\max}}>1$, with an additional
logarithmic factor $O(N\log N)$ for W- and K-cycles when
$d={N_{\max}}=1$ (Supplementary Section~S2.3). Level
matrices and ILU(0) factors are reused while $\sigma$ is unchanged, and
existing diagonal blocks are preserved when adaptive FSP increases
${N_{\max}}$ (Supplementary Section~S3.2). Optional
reduction under reflection symmetry is described in Supplementary
Sections~S4.3 and~S4.4.

\subsection{Algorithms}\label{sec:algorithms}
Algorithm~\ref{alg:time} summarizes the implicit time integration loop,
Algorithm~\ref{alg:method} specifies the recursive multigrid cycle, and
Algorithms~\ref{alg:transfer-setup}--\ref{alg:prolongation} give the
construction and application of the transfer operators.

The implementation stores two arrays of length $N_l$ on each level rather
than sparse transfer matrices. For fine state $X^{(i)}$,
\texttt{parent[i]} records the coarse index $j$ of ${X^c}(X^{(i)})$, and
\texttt{weight[i]} stores its multinomial weight $P_{ij}$, encoding
\begin{equation}\label{eq:stored-transfers}
 R_{\texttt{parent[i]},i}=1,\qquad
 P_{i,\texttt{parent[i]}}=\texttt{weight[i]},
\end{equation}
with all other entries zero. Algorithm~\ref{alg:transfer-setup} builds
these arrays once during setup, evaluating factorials through their
logarithms and normalizing each column of $P$ to unity. While the
permutation in \eqref{eq:transfer-blocks} illustrates the block
structure, the solver retains its population and lexicographic voxel
ordering on every level, as shown in Supplementary
Figures~S1 and~S2.

\begin{algorithm}[!htbp]
\caption{Advance the RDME probability distribution in time}
\label{alg:time}
\begin{algorithmic}[1]
\STATE Given the retained states, $\boldsymbol p(0)$, final time $T$,
step lengths, time integration method, cycle type, and tolerance $\varepsilon$.
\STATE Construct the voxel hierarchy down to one voxel and its transfers
by Algorithm~\ref{alg:transfer-setup}, and form the generators
$\mathbf A_l$ using \eqref{eq:coarse-generator}.
\STATE Initialize $t\gets0$ and $\boldsymbol p\gets\boldsymbol p(0)$.
\FOR{each prescribed step of length $\Delta t$ up to $T$}
 \STATE Form $\sigma$ and $b$ from $\boldsymbol p$ using \eqref{eq:time-steps}.
 \STATE Form $\mathbf A_{\sigma,l}=\sigma I-\mathbf A_l$, relaxation factors,
 and the bottom factorization, reusing them when $\sigma$ is unchanged.
 \STATE Initialize $x\gets\boldsymbol p$ and $r\gets b-\mathbf A_{\sigma,0}x$.
 \WHILE{$\|r\|_1/\sigma>\varepsilon$}
  \STATE $x\gets x+\Cycle(0,r)$ using Algorithm~\ref{alg:method}.
  \STATE $r\gets b-\mathbf A_{\sigma,0}x$.
 \ENDWHILE
 \STATE Accept $\boldsymbol p\gets x$ and advance $t\gets t+\Delta t$.
\ENDFOR
\RETURN the fine probability distribution $\boldsymbol p$ at $T$.
\end{algorithmic}
\end{algorithm}

\begin{algorithm}[!htbp]
\caption{One multigrid cycle $\Cycle( l ,b)$}
\label{alg:method}
\begin{algorithmic}[1]
\STATE Input level $ l $ and right-hand side $b$, and initialize $x\gets0$.
\IF{$ l ={N_{\mathrm l}}$ (one-voxel CME)}
 \RETURN $\mathbf A_{\sigma,N_{\mathrm l}}^{-1}b$ by the bottom direct solve.
\ENDIF
\STATE Apply forward relaxation to $\mathbf A_{\sigma,l} x=b$.
\STATE Form $r\gets b-\mathbf A_{\sigma,l} x$ and restrict $b_c\gets R_l r$
using Algorithm~\ref{alg:restriction}.
\STATE Recurse toward the CME to compute $e_c$ by \eqref{eq:vw}
for the V- or W-cycle, or by residual minimization for the K-cycle.
\STATE Prolong and add $x\gets x+P_l e_c$ using Algorithm~\ref{alg:prolongation}.
\STATE Apply reverse relaxation to $\mathbf A_{\sigma,l} x=b$ and return $x$.
\end{algorithmic}
\end{algorithm}

\begin{algorithm}[!htbp]
\caption{Construct restriction and prolongation for adjacent levels}
\label{alg:transfer-setup}
\begin{algorithmic}[1]
\STATE Given fine and coarse state lists, aggregates $C_k$, and volumes,
initialize \texttt{sum[j]}$\gets0$ for every coarse index $j$.
\FOR{each fine state $X$ with index $i$}
 \STATE ${X^c}_{k,s}\gets\sum_{v\in C_k}X_{v,s}$ for every aggregate and species.
 \STATE Let $j$ be the index of ${X^c}$ in the coarse state list, with \texttt{parent[i]}$\gets j$.
 \STATE $\displaystyle\texttt{weight[i]}\gets
  \prod_{k,s}\left[\frac{{X^c}_{k,s}!}{\prod_{v\in C_k}X_{v,s}!}
  \prod_{v\in C_k}\left(\frac{\omega_v}{\Omega_k}\right)^{X_{v,s}}\right]$
 \STATE \texttt{sum[j]}$\gets$\texttt{sum[j]}$+$\texttt{weight[i]}.
\ENDFOR
\FOR{each fine index $i$}
 \STATE \texttt{weight[i]}$\gets$\texttt{weight[i]}$/\texttt{sum[parent[i]]}$.
\ENDFOR
\RETURN \texttt{parent} and \texttt{weight}.
\end{algorithmic}
\end{algorithm}

\begin{algorithm}[!htbp]
\caption{Restrict a fine residual}
\label{alg:restriction}
\begin{algorithmic}[1]
\STATE Given fine residual $r$ and \texttt{parent}, initialize $b_c\gets0$.
\FOR{each fine index $i$}
 \STATE $(b_c)_{\texttt{parent[i]}}\gets(b_c)_{\texttt{parent[i]}}+r_i$.
\ENDFOR
\RETURN $b_c=Rr$.
\end{algorithmic}
\end{algorithm}

\begin{algorithm}[!htbp]
\caption{Prolong and add a coarse correction}
\label{alg:prolongation}
\begin{algorithmic}[1]
\STATE Given coarse correction $e_c$, fine iterate $x$,
\texttt{parent}, and \texttt{weight}.
\FOR{each fine index $i$}
 \STATE $x_i\gets x_i+\texttt{weight[i]}(e_c)_{\texttt{parent[i]}}$.
\ENDFOR
\RETURN the updated fine iterate $x$.
\end{algorithmic}
\end{algorithm}

\FloatBarrier
\section{Approximation Properties}\label{sec:analysis}
We first state the Galerkin closure property verifying that the coarse
reaction and diffusion rates in \eqref{eq:coarse-diffusion} and
\eqref{eq:coarse-reaction} reproduce $\mathbf A_c=R\mathbf AP$. We then
characterize the fine error omitted by the coarse projection $PR$ under
reversible diffusion, showing that $PR$ averages the relative error over
each coarse class and that the omitted variance is bounded by the
internal aggregate Dirichlet energy with a constant independent of
population. Finally, we state the $\ell_1$ resolvent bound used by the
stopping rule. Complete proofs of Theorem~\ref{thm:reaction},
Lemma~\ref{lem:averaging}, Proposition~\ref{prop:aggregate}, and
Proposition~\ref{prop:residual} are given in Appendix~\ref{app:proofs}.

\subsection{Coarse Generators}\label{sec:coarse-properties}
For distinct coarse states $j\ne k$, the off-diagonal sum in
\eqref{eq:coarse-entry} is nonnegative because the two fine classes are
disjoint and $P_{ik}\ge0$. Moreover, \eqref{eq:transfer-identities} and
$\one^\top\mathbf A\le0$ imply $\one^\top\mathbf A_c=\one^\top\mathbf AP\le0$,
so $\mathbf A_c$ is a Markov generator with absorbing boundary losses.
Averaging the diffusive jumps of each molecule under the binomial
marginal of \eqref{eq:P} yields the coarse jump rate
\eqref{eq:coarse-diffusion}, and evaluating the factorial moments of
\eqref{eq:P} establishes the coarse reaction propensities.

\begin{theorem}\label{thm:reaction}
If $\mathcal X$ contains complete population blocks and $c_r$ is
uniform within each aggregate, then the reaction part of
$\mathbf A_c=R\mathbf AP$ has propensity \eqref{eq:coarse-reaction},
including losses to omitted destination states.
\end{theorem}

\subsection{Diffusion and the Coarse Approximation}\label{sec:spectrum}\label{sec:approximation}
We now construct the equilibrium weights $\pi_i$ and derive the graph
energy introduced in \eqref{eq:diffusion-energy}. Assume that the voxel
graph is connected and that every edge has positive diffusive jump rates
satisfying $\omega_vd_{vw}^{(s)}=\omega_wd_{wv}^{(s)}$. Because diffusion
preserves each species total, every population block is a finite closed
communicating class of $\mathbf A_{\mathrm d}$. For one species at total
population $k$, the multinomial distribution
\begin{equation}\label{eq:equilibrium}
 \pi(X)=\frac{k!}{\prod_vX_v!}
          \prod_v\left(\frac{\omega_v}{\Omega}\right)^{X_v},
 \qquad\Omega=\sum_v\omega_v,
\end{equation}
assigns positive weight to each arrangement and sums to one by the
multinomial theorem. For a jump $X\to X'$ moving one molecule from voxel
$v$ to voxel $w$, the weights satisfy
\[
 \frac{\pi(X')}{\pi(X)}
 =\frac{X_v}{X_w+1}\frac{\omega_w}{\omega_v},\qquad
 \pi(X)d_{vw}X_v=\pi(X')d_{wv}(X_w+1),
\]
equating the forward and reverse equilibrium fluxes across every edge and
implying stationarity under $\mathbf A_{\mathrm d}$. For several species,
the product of their single-species distributions \eqref{eq:equilibrium}
gives the normalized diffusion equilibrium on each population block,
which may be scaled by any positive block weight $m(k)>0$ without
requiring an equilibrium assumption for the reaction channels.

Letting $\pi_i=\pi(X^{(i)})$ and $\Pi=\operatorname{diag}(\pi)$, detailed
balance and stationarity read
\[
 \pi_i(\mathbf A_{\mathrm d})_{ji}=\pi_j(\mathbf A_{\mathrm d})_{ij},
 \qquad \mathbf A_{\mathrm d}\pi=0,
\]
so multiplying $\mathbf A_{\mathrm d}$ on the right by $\Pi$ produces a
symmetric positive semidefinite graph Laplacian~\cite{aldous2002,chung1997},
\begin{equation}\label{eq:conductance}
 \mathbf L=-\mathbf A_{\mathrm d}\Pi,
\end{equation}
with nonpositive off-diagonal entries and zero row sums. Writing an error
vector $e$ in terms of its relative entries $f_i=e_i/\pi_i$ as $e=\Pi f$,
pure diffusion $\dot e=\mathbf A_{\mathrm d}e$ takes the form
\[
 \dot f_i=\sum_{j\ne i}(\mathbf A_{\mathrm d})_{ji}(f_j-f_i),
 \qquad \Pi\dot f=-\mathbf Lf,
\]
and differentiating $\|e\|_{\pi^{-1}}^2=f^\top\Pi f$ yields the energy
dissipation identity
\[
 \frac12\frac{d}{dt}\|e\|_{\pi^{-1}}^2
 =-f^\top\mathbf Lf
 =-\frac12\sum_{i\ne j}\pi_i(\mathbf A_{\mathrm d})_{ji}
                         (f_i-f_j)^2,
\]
matching the graph energy in \eqref{eq:diffusion-energy}. If
$\mathbf L\phi=\lambda\Pi\phi$, then $\mathbf A_{\mathrm d}\Pi\phi=-\lambda\Pi\phi$
and the mode $\Pi\phi$ decays as $e^{-\lambda t}$ with decay rate
$\lambda=(\phi^\top\mathbf L\phi)/(\phi^\top\Pi\phi)$.

\begin{figure}[!htbp]
\centering
\begin{tikzpicture}[x=1cm,y=1cm,>=Stealth,font=\small,
 st/.style={draw,fill=white,rounded corners=1.5pt,inner sep=2pt,font=\scriptsize},
 vox/.style={draw,minimum width=.62cm,minimum height=.55cm},
 inag/.style={very thick},
 crossag/.style={densely dashed,gray}]
% (a) voxel graph: chain v1-v2-v3 with aggregates {v1,v2},{v3}
\node[anchor=west] at (-.2,3.55) {(a) Voxels and aggregates};
\node[vox] (v1) at (0.3,2.6) {$v_1$};
\node[vox] (v2) at (1.3,2.6) {$v_2$};
\node[vox] (v3) at (2.6,2.6) {$v_3$};
\draw[inag] (v1)--(v2);
\draw[crossag] (v2)--(v3);
\draw[rounded corners=3pt,thick] (-.15,2.15) rectangle (1.75,3.05);
\draw[rounded corners=3pt,thick] (2.15,2.15) rectangle (3.05,3.05);
\node[font=\footnotesize] at (0.8,1.85) {$C_1$};
\node[font=\footnotesize] at (2.6,1.85) {$C_2$};
\node[font=\scriptsize,align=left,anchor=north west] at (-.2,1.35)
 {\tikz\draw[inag](0,0)--(.45,0); jump inside an aggregate\\[2pt]
  \tikz\draw[crossag](0,0)--(.45,0); jump between aggregates};
% (b) state graph at k=3, rows = coarse classes N=(N_1,N_2)
\begin{scope}[xshift=7.4cm]
\node[anchor=west] at (-3.0,3.55) {(b) States with $k=3$, grouped by ${X^c}=({X^c}_1,{X^c}_2)$};
\def\s{1.25}\def\h{1.05}
\foreach \c in {0,...,3}{
 \pgfmathsetmacro{\w}{(3-\c)*\s/2+.72}
 \fill[gray!12,rounded corners=4pt] (-\w,\c*\h-.32) rectangle (\w,\c*\h+.32);
 \pgfmathtruncatemacro{\Na}{3-\c}
 \node[font=\scriptsize,anchor=west] at (2.75,\c*\h) {${X^c}=(\Na,\c)$};
}
\foreach \c in {0,...,3}{
 \pgfmathtruncatemacro{\m}{3-\c}
 \foreach \b in {0,...,\m}{
  \pgfmathtruncatemacro{\a}{\m-\b}
  \node[st] (n\a\b\c) at ({(\b-\a)*\s/2},{\c*\h}) {$(\a,\b,\c)$};
 }}
% v1<->v2 jumps: same row neighbours (inside C_1)
\foreach \c in {0,...,2}{
 \pgfmathtruncatemacro{\m}{2-\c}
 \foreach \b in {0,...,\m}{
  \pgfmathtruncatemacro{\a}{\m-\b+1}\pgfmathtruncatemacro{\aa}{\a-1}\pgfmathtruncatemacro{\bb}{\b+1}
  \draw[inag] (n\a\b\c)--(n\aa\bb\c);
 }}
% v2<->v3 jumps: (a,b,c)->(a,b-1,c+1), between rows
\foreach \c in {0,...,2}{
 \pgfmathtruncatemacro{\m}{3-\c}
 \foreach \b in {1,...,\m}{
  \pgfmathtruncatemacro{\a}{\m-\b}\pgfmathtruncatemacro{\bb}{\b-1}\pgfmathtruncatemacro{\cc}{\c+1}
  \draw[crossag] (n\a\b\c)--(n\a\bb\cc);
 }}
% prolongation weights for the class N=(3,0)
\foreach \a/\b/\wt in {3/0/\tfrac18,2/1/\tfrac38,1/2/\tfrac38,0/3/\tfrac18}
 \node[font=\scriptsize] at ({(\b-\a)*\s/2},-.55) {$\wt$};
\node[font=\scriptsize,anchor=east] at (-2.65,-.55) {weights of $P$:};
\end{scope}
\end{tikzpicture}
\caption{Diffusion state graph for one species with $k=3$ on a chain
of three voxels of equal volume, with aggregates $C_1=\{v_1,v_2\}$
and $C_2=\{v_3\}$. Each shaded row is one coarse state ${X^c}$, and $PR$
averages the relative error over the fine states in that row.
Solid edges represent jumps inside an aggregate and contribute to
$\mathcal E_{\mathrm{in}}$, whereas dashed edges change the coarse
state. The weights below the lowest row are its prolongation weights.}
\label{fig:state-classes}
\end{figure}
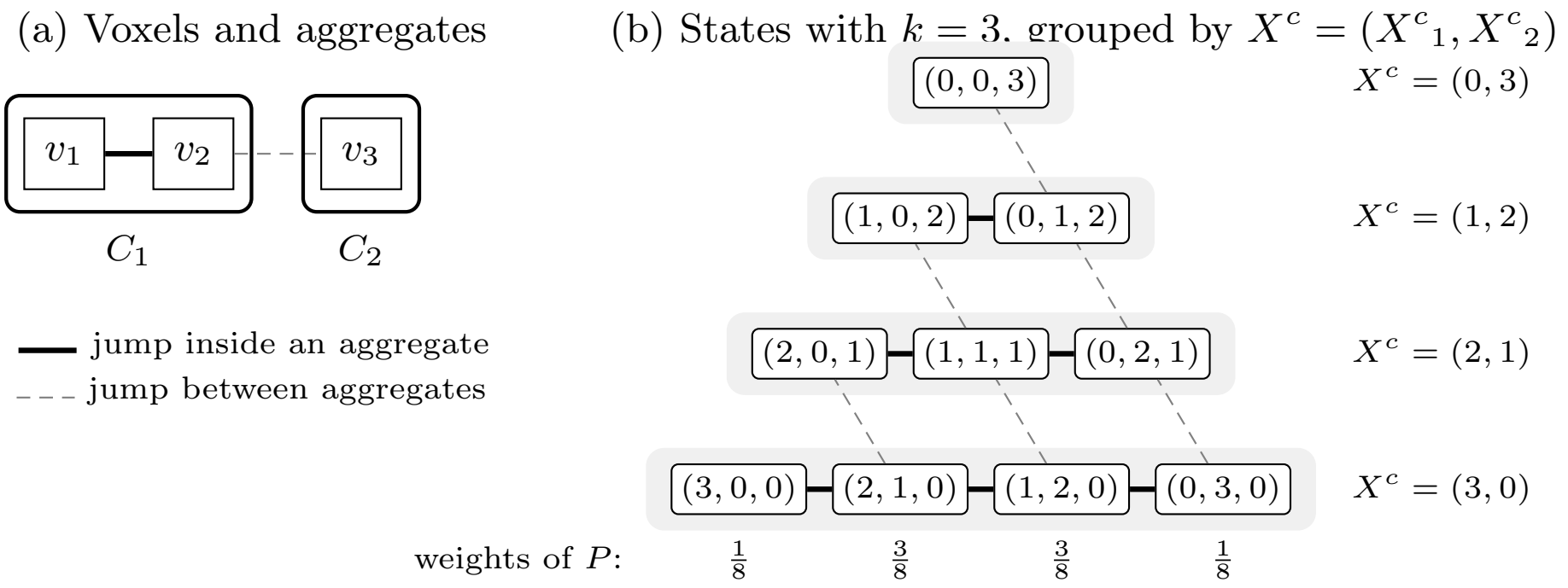

Conditioning the global equilibrium weights $\pi_i$ on the coarse
aggregate counts recovers the multinomial prolongation weights
\eqref{eq:P} and characterizes the projection error $e-PRe$ as a
conditional variance.
\begin{lemma}\label{lem:averaging}
For reversible diffusion,
\begin{equation}\label{eq:conditional-transfer}
 P_{ij}=\frac{\pi_i}{(R\pi)_j}\quad\text{when }R_{ji}=1,
\end{equation}
and the error omitted by $PR$ satisfies
\begin{equation}\label{eq:variance}
 \|e-PRe\|_{\pi^{-1}}^2
 =\sum_j\ \sum_{i:\,R_{ji}=1}\pi_i
       \left[f_i-\frac{(Re)_j}{(R\pi)_j}\right]^2.
\end{equation}
\end{lemma}
For the represented error $PRe$, the relative entries $(PRe)_i/\pi_i$
equal the conditional average $(Re)_j/(R\pi)_j$ of $f_i=e_i/\pi_i$ over
all fine states in coarse class $j$, and \eqref{eq:variance} measures the
variance lost in this averaging, as illustrated in
Figure~\ref{fig:state-classes}.

Diffusive jumps that preserve the coarse state move molecules only inside
aggregates. Let $\mathcal E_{\mathrm{in}}(f)$ denote the sum in
\eqref{eq:diffusion-energy} restricted to jumps inside aggregates. For
each species and each aggregate containing at least two voxels, consider
a single molecule confined to that aggregate, and let $\lambda_*>0$ be
the minimum of the smallest positive eigenvalues of these single-molecule
negative diffusion generators across all species and multi-voxel
aggregates.
\begin{proposition}\label{prop:aggregate}
Under these assumptions,
\[
 \|e-PRe\|_{\pi^{-1}}^2
 \le\lambda_*^{-1}\mathcal E_{\mathrm{in}}(f).
\]
\end{proposition}
Because $\lambda_*$ is determined by single-molecule diffusion on each
aggregate, the constant $\lambda_*^{-1}$ depends on the aggregates and
jump rates but is independent of the molecular population. Relative
errors that vary little across diffusive jumps inside aggregates are
therefore approximated by the coarse space with accuracy independent of
population size. A convergence estimate for the complete cycle would also
require bounds for relaxation and recursive coarse solves. When the full
reacting system satisfies detailed balance, exact coarse correction also
minimizes an operator energy norm, as shown in Supplementary
Section~S2.2.

\subsection{Accuracy of the Linear Solves}\label{sec:algebraic-error}
The stopping rule bounds the $\ell_1$ algebraic error in a completed
linear solve through the computable fine residual, without requiring
reversibility or an equilibrium assumption.
\begin{proposition}\label{prop:residual}
If $\mathbf A$ has nonnegative off-diagonal entries and nonpositive
column sums, then for $\mathbf A_\sigma =\sigma I-\mathbf A$ with
$\sigma>0$,
\begin{equation}\label{eq:residual}
 \|x-\mathbf A_\sigma^{-1}b\|_1
 \le\sigma^{-1}\|b-\mathbf A_\sigma x\|_1.
\end{equation}
\end{proposition}
When $\|b-\mathbf A_\sigma x\|_1/\sigma\le\varepsilon$, any marginal
probability formed by summing a subset of entries of $x$ has algebraic
error at most $\varepsilon$. Errors from earlier steps, time
discretization, and FSP truncation are assessed separately.

\FloatBarrier
\section{Numerical Experiments}\label{sec:experiments}
The experiments first examine how spatial levels, cycle choice, and
relaxation affect convergence as the population cap and spatial grid are
refined, comparing individual linear solves at a common tolerance through
\eqref{eq:residual}. We then apply the solver to transient distributions
with two population peaks and to rare switching probabilities, assessing
population truncation and time discretization separately.

\subsection{Experimental Setup}
We test single-species models on uniform Cartesian grids of total volume
one with reflecting boundaries. Let $X_v$ be the count in voxel $v$ and
$k=\sum_vX_v$ the total population, whose marginal distribution is
obtained by summing probabilities over all states with total $k$.

We consider a linear immigration-degradation model with birth propensity
$2\omega_v$ and death propensity $X_v$, together with the bistable
Schl\"ogl system~\cite{schlogl1972}, whose birth and death propensities in
voxel $v$ are
\[
 \begin{aligned}
 a_{+,v}(X_v)&=c_3\omega_v+\frac{c_1}{\omega_v}\binom{X_v}{2},\\
 a_{-,v}(X_v)&=c_4X_v+\frac{c_2}{\omega_v^2}\binom{X_v}{3},\\
 (c_1,c_2,c_3,c_4)&=10^{-4}(133,6,6924.125,1147.25).
 \end{aligned}
\]
For individual linear solves, we let $D=1$ and $b=\sigma\boldsymbol p(0)$
and initialize the iteration with $x=\boldsymbol p(0)$, corresponding to
one backward Euler step. The linear model uses $\sigma=0.5$ and starts
with one molecule in each of the first two voxels. The Schl\"ogl tests
use four voxels on a chain of length one, $\sigma=0.05$, and two
molecules per voxel. Transient experiments specify their own initial
states and step sizes.

The solver uses double precision and compressed sparse row (CSR) matrices
on an Apple M1 Pro with 16 GB of memory. Unless stated otherwise, the
stopping tolerance in \eqref{eq:residual} is $\varepsilon=10^{-9}$. Point
relaxation uses two forward and two reverse sweeps per level and block
relaxation uses one sweep in each direction, except in direct comparisons
where both use a single forward and reverse sweep. The ILU-GMRES baseline
applies ILU(0) right preconditioning to the full matrix $\mathbf A_\sigma$
in GMRES(30) with the same stopping bound \eqref{eq:residual}, and runs
reaching the iteration limit are marked as unconverged. Reported timings
come from single runs and include setup, which is reused across steps
when $\sigma$ is unchanged. Repeated-run implementation timings are
reported in Supplementary Section~S4.4.

\subsection{Convergence and Refinement}\label{sec:cycle-results}
\begin{figure}[!htbp]
\centering\includegraphics[width=\linewidth]{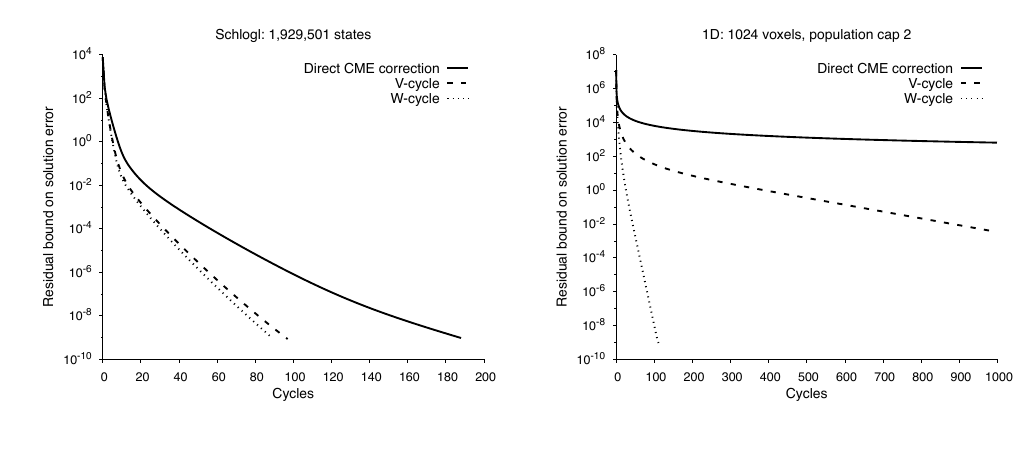}
\caption{Residual bound \eqref{eq:residual} for the Schl\"ogl problem on four
voxels at cap 80 (left) and the linear model on 1,024 voxels at cap 2
(right), with the common point relaxation and parameters specified
in the setup. Direct CME correction omits intermediate spatial
levels, which V and W retain. Table~\ref{tab:cycles} gives the
K-cycle results for the problem on the right.}
\label{fig:convergence}
\end{figure}

We first examine the effect of retaining an intermediate spatial level on
the four-voxel Schl\"ogl problem at cap $80$, where the fine level has
$1{,}929{,}501$ states, the two-voxel level has $3{,}403$ states, and the
bottom CME has $81$ states. With identical point relaxation and initial
guess, omitting the two-voxel level requires $188$ iterations, whereas
the $4\to2\to1$ hierarchy requires $97$ V-cycles or $89$ W-cycles
(Figure~\ref{fig:convergence}, left).

On the linear problem with $1{,}024$ voxels at cap $2$ ($11$ levels and
$525{,}825$ fine states), the V-cycle fails to converge within $1{,}000$
cycles, whereas the W- and K-cycles converge in $111$ and $55$ cycles
(Figure~\ref{fig:convergence}, right, and Table~\ref{tab:cycles}). The
K-cycle nearly halves the solve time relative to the W-cycle, while their
shared setup cost reduces the saving in total time to about $17\%$.

\begin{table}[!htbp]
\centering
\caption{Cycles for the linear problem on 1,024 voxels at cap 2, $D=1$, and
$\sigma=0.5$, using the same levels and point relaxation. V reaches
the iteration limit, whereas W and K meet tolerance $10^{-9}$.
Times are in seconds and peak memory is in GiB, with one run per cycle type.}
\label{tab:cycles}
\begin{tabular}{lrrrrr}
\toprule
cycle & iterations & final bound & setup & solve & GiB\\
\midrule
V & 1000 (limit) & $3.36\times10^{-3}$ & 13.33 & 42.69 & 0.214\\
W & 111 & $9.33\times10^{-10}$ & 13.36 & 6.92 & 0.228\\
K & 55 & $8.87\times10^{-10}$ & 13.47 & 3.35 & 0.223\\
\bottomrule
\end{tabular}
\end{table}

\phantomsection\label{sec:spatial-results}
To test population refinement, we increase ${N_{\max}}$ from $2$ to $8$
on $16$ voxels while retaining the five spatial levels. In
Table~\ref{tab:population}, a $4{,}800$-fold increase in the state count
raises the W-cycle count only from $30$ to $37$, with K-cycle counts of
$30$, $33$, $35$, and $35$, whereas direct CME correction rises from
$214$ to $501$ iterations. These nearly flat multilevel counts are
consistent with the population-independent constant in
Proposition~\ref{prop:aggregate}, although that approximation bound alone
does not establish the full cycle convergence rate.

\begin{table}[!htbp]
\centering
\caption{Population refinement on 16 voxels for the linear model, $D=1$,
$\sigma=0.5$, with default point relaxation and tolerance $10^{-9}$.
CME correction omits intermediate spatial levels.}
\label{tab:population}
\begin{tabular}{rrrr}
\toprule
cap & states & CME correction & W-cycle\\
\midrule
% Source: paper/data/full_multigrid/graph_gap_cap{2,4,6,8}_{direct,direct1200,w,kcycle}.csv
2 & 153     & 214 & 30\\
4 & 4,845   & 359 & 34\\
6 & 74,613  & 455 & 36\\
8 & 735,471 & 501 & 37\\
\bottomrule
\end{tabular}
\end{table}

\begin{figure}[!htbp]
\centering\includegraphics[width=\linewidth]{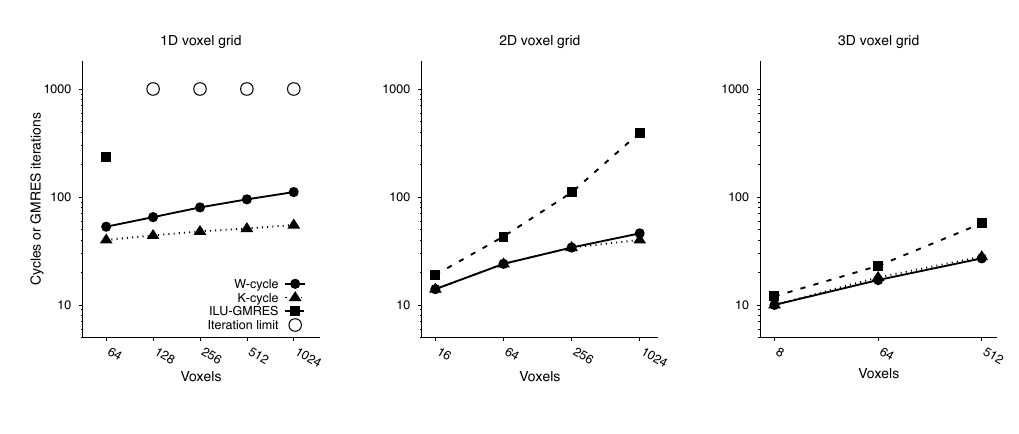}
\caption{Spatial refinement for the linear model at cap 2, with W and K
reporting cycles and ILU-GMRES reporting iterations. All methods
use tolerance $10^{-9}$, and the cycles use two point Gauss--Seidel
sweeps in each direction. Open markers indicate runs that reach the iteration limit. The
counts compare convergence, with different costs per iteration for
the three methods.}
\label{fig:scaling}
\end{figure}

For spatial refinement we fix ${N_{\max}}=2$ and increase the voxel count
in one, two, and three dimensions. This small cap permits grids up to
$1{,}024$ voxels ($525{,}825$ states) to isolate linear-solver scaling on
the specified finite matrices, while FSP truncation accuracy is examined
in the transient experiments below. Figure~\ref{fig:scaling} compares the
W- and K-cycles against ILU-GMRES. In one dimension, refining from $64$
to $1{,}024$ voxels raises the W-cycle count from $53$ to $111$ and the
K-cycle count through $40$, $44$, $48$, $51$, and $55$, whereas ILU-GMRES
converges within $1{,}000$ iterations only on $64$ voxels (and remains at
bound $0.207$ after $5{,}000$ iterations on $1{,}024$ voxels).

The largest two-dimensional grid ($32\times32$) requires $46$ W-cycles,
$40$ K-cycles, or $394$ ILU-GMRES iterations, compared with $27$, $28$,
and $57$ on the largest three-dimensional grid ($8\times8\times8$). The
K-cycle is therefore most advantageous on the deepest hierarchy in one
dimension and does not reduce the cycle count on every grid.

\subsection{Smoothing and Computational Cost}
Table~\ref{tab:block-smoothing} compares point and block Gauss--Seidel
relaxation using one forward and reverse sweep per level. Block updates
reduce both the W-cycle count and total elapsed time across all tested
problems. For the four-voxel Schl\"ogl benchmark at cap $40$, switching
from point to block smoothing reduces the cycle count from $110$ to $40$
and cuts total time from $0.706$ to $0.362$ seconds.

\begin{table}[!htbp]
\centering
\caption{Point and block relaxation in W-cycles for one linear solve,
using one forward and one reverse sweep per level. Each pair of
entries lists the point result first and the block result second.
Times include setup and solve, in seconds from individual runs.}
\label{tab:block-smoothing}
\begin{tabular}{lrrrrr}
\toprule
model & voxels & cap & states & cycles & total time (s)\\
\midrule
Schl\"ogl & 4 & 40 & 135,751 & 110 / 40 & 0.706 / 0.362\\
Schl\"ogl & 4 & 60 & 635,376 & 154 / 56 & 4.884 / 2.343\\
linear & 16 & 8 & 735,471 & 50 / 28 & 2.888 / 2.265\\
linear & 64 & 4 & 814,385 & 67 / 48 & 4.590 / 4.390\\
\bottomrule
\end{tabular}
\end{table}

Block relaxation alone does not replace coarse correction. After $300$
forward and reverse block sweeps without coarse correction on the cap-$40$
Schl\"ogl problem, the residual bound remains above $0.248$. With the
$4\to2\to1$ hierarchy, W- and K-cycles require $40$ and $41$ cycles at
cap $40$, and $56$ and $57$ cycles at cap $60$. On these shallow
hierarchies, strengthening the smoother from point to block relaxation
has a much larger effect than changing the W-cycle to a K-cycle.

To compare multigrid with sparse direct factorization, we test
KLU~\cite{davis2010} with approximate minimum degree (AMD)
ordering~\cite{amestoy1996}. On the $1{,}024$-voxel linear benchmark, the
point W-cycle takes $20.3$ seconds including setup and uses $0.23$ GiB of
memory, whereas KLU takes $31.4$ seconds and $1.16$ GiB
(Table~\ref{tab:cost-storage}). Once factored, however, KLU solves each
subsequent right-hand side in $0.19$ seconds compared with $6.92$ seconds
for the W-cycle, recovering its larger setup cost after about three
solves when the factors fit in memory.

\begin{table}[!htbp]
\centering
\caption{Cost at tolerance $10^{-9}$, with times in seconds and peak memory
in GiB. The linear problem has 1,024 voxels and $\sigma=0.5$, while
the Schl\"ogl transients on four voxels use $T=100$, $\Delta t=5$,
and 21 solves with reused setup. All use $D=1$. Point and block W
use their default numbers of sweeps, and a dash denotes direct solution.}
\label{tab:cost-storage}
\begingroup\small\setlength{\tabcolsep}{4pt}
\begin{tabular}{lrrrrrr}
\toprule
method & ${N_{\max}}$ & states & cycles & setup & solve & GiB\\
\midrule
\multicolumn{7}{l}{Linear model}\\
% Source: paper/data/full_multigrid/klu_1024_voxels.csv
KLU & 2 & 525,825 & --- & 31.17 & 0.19 & 1.16\\
% Source: paper/data/full_multigrid/history_spatial_w_solve.csv
point W & 2 & 525,825 & 111 & 13.36 & 6.92 & 0.23\\
\midrule
\multicolumn{7}{l}{Schl\"ogl transient}\\
% Source: paper/data/full_multigrid/transient_k96_s20.csv
point W & 96 & 3,921,225 & 1481 & 1.06 & 435.57 & 1.02\\
% Source: paper/data/full_multigrid/bistability_block_k96_s20.csv
block W & 96 & 3,921,225 & 1072 & 1.56 & 261.47 & 1.50\\
% Source: paper/data/full_multigrid/transient_k112_s20.csv
point W & 112 & 7,160,245 & 1481 & 1.74 & 816.72 & 1.71\\
\bottomrule
\end{tabular}
\endgroup
\end{table}

On the nonlinear Schl\"ogl system on four voxels, KLU factors the matrix
at cap $20$ but exceeds the $2$ GiB SuiteSparse allocation limit at cap
$32$ and higher caps, where symbolic analysis predicts $1.2\times10^8$
nonzeros per triangular factor at cap $32$ and $5.1\times10^{10}$ at cap
$80$. Multigrid completes the transient integrations at caps $96$ and
$112$ within $1.71$ GiB while reusing its setup across all $21$ steps.

\subsection{Transient Distributions}\label{sec:applications}
\begin{figure}[!htbp]
\centering\includegraphics[width=\linewidth]{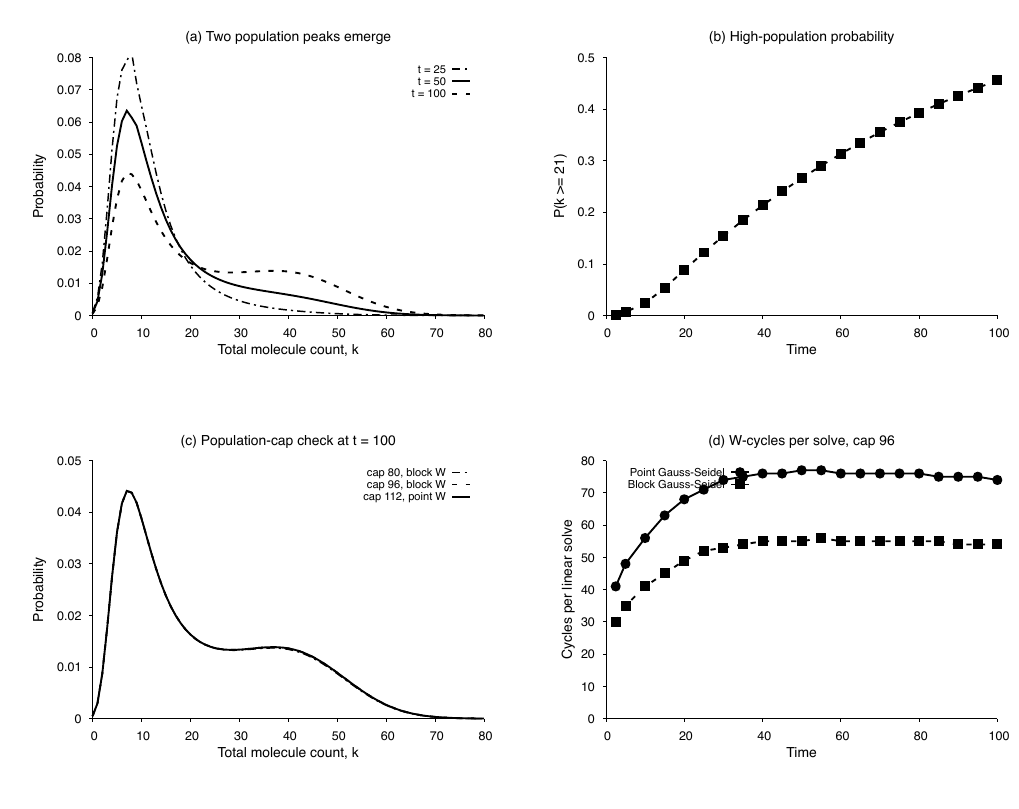}
\caption{Schl\"ogl evolution on four voxels from $(2,2,2,2)$, $D=1$,
$\Delta t=5$, and algebraic tolerance $10^{-9}$. Panels (a,b)
use cap 96 and block W. Panel (c) compares population caps at
$T=100$, and panel (d) compares the number of point and block
W-cycles per solve at cap 96.}
\label{fig:transient}
\end{figure}

We next integrate the Schl\"ogl distribution from $(2,2,2,2)$ to $T=100$
on four voxels with $D=1$ using Crank--Nicolson with two initial backward
Euler half-steps, without clipping negative entries or renormalizing the
solution vector. Figure~\ref{fig:transient} shows the evolution of the
population marginal, the mass $\mathbb P(k\ge21)$ above $k=20$, and the
W-cycle counts per solve.

\begin{table}[!htbp]
\centering
\caption{Population refinement for the Schl\"ogl transient on four voxels at
$T=100$, $\Delta t=5$, with W and tolerance $10^{-9}$. The reported mass is subject to errors from time discretization
and linear solves.}
\label{tab:cap}
\begin{tabular}{lrrrr}
\toprule
relaxation & ${N_{\max}}$ & states & retained mass & $\mathbb P(k\geq21)$\\
\midrule
block & 80  & 1,929,501 & 0.9932912374 & 0.4491884933\\
block & 96  & 3,921,225 & 0.9999815048 & 0.4556061510\\
point & 112 & 7,160,245 & 0.9999999950 & 0.4556239278\\
\bottomrule
\end{tabular}
\end{table}

At $T=100$, the distribution exhibits two population modes at $k=7$ and
$k=37$. Increasing the cap from $80$ to $96$ changes $\mathbb P(k\ge21)$
by $6.42\times10^{-3}$, and increasing it from $96$ to $112$
($7{,}160{,}245$ fine states and $113$ CME states) changes
$\mathbb P(k\ge21)$ by $1.78\times10^{-5}$ and the population marginal by
$1.85\times10^{-5}$ in $\ell_1$ (Table~\ref{tab:cap}). At cap $96$, point
and block relaxation produce final marginals agreeing to
$8.15\times10^{-13}$ in $\ell_1$, while block relaxation reduces runtime
from $437$ to $263$ seconds and increases peak memory from $1.02$ to
$1.50$ GiB.

At cap $112$, halving the step size through $\Delta t=5$, $2.5$, and
$1.25$ gives $\mathbb P(k\ge21)=0.4556239278$, $0.4557196041$, and
$0.4557434892$ in $818$, $1{,}716$, and $2{,}438$ seconds, respectively.
The successive differences $9.57\times10^{-5}$ and $2.39\times10^{-5}$
decrease by a factor of four, consistent with second-order temporal
convergence, and Supplementary Section~S3.1 provides
an independent uniformization check on two voxels.

\subsection{Rare Population Switching}\label{sec:rare-switching}
\begin{figure}[!htbp]
\centering
\includegraphics[width=\linewidth]{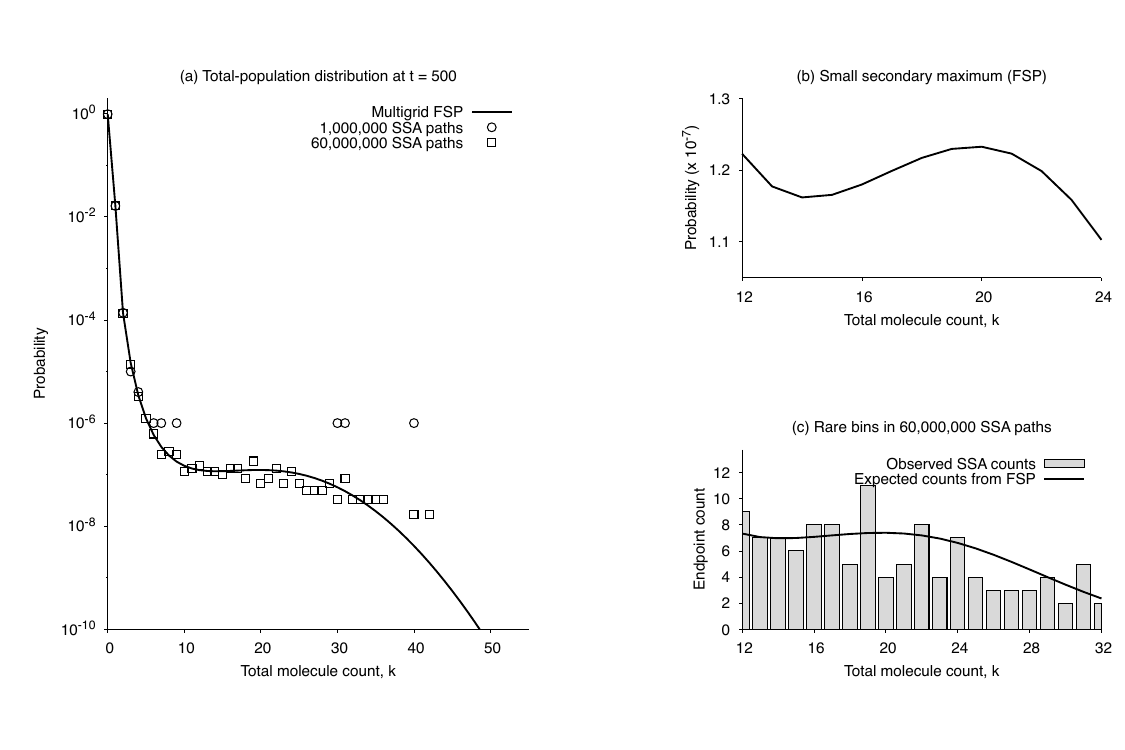}
\caption{Rare switching at $T=500$. Panel (a) compares FSP at cap 72,
$\Delta t=5$, with SSA, omitting bins with zero observed endpoints
from the logarithmic axis. The first million paths belong to the
sample of 60 million paths. Panels (b,c) show the secondary peak
and the observed and expected counts for individual integer
populations. Supplementary Table~S1
gives the FSP refinements.}
\label{fig:rare-switching}
\end{figure}

To test direct resolution of rare events~\cite{munsky2006}, we initialize
the four-voxel Schl\"ogl chain in the empty state $(0,0,0,0)$ with $D=1$
and $(c_1,c_2,c_3,c_4)=(0.0258,0.0024,0.00138,0.0829)$, and integrate to
$T=500$. The resulting population marginal has a dominant mode at $k=0$
and a faint secondary peak at $k=20$ separated by a minimum near $k=14$
(Figure~\ref{fig:rare-switching}). At cap $72$ and $\Delta t=5$ with
tolerance $10^{-14}$ ($1{,}282{,}975$ states), the multigrid FSP solver
gives $\mathbb P(k(T)\ge16)=1.7746\times10^{-6}$ and
$\mathbb P(k(T)=20)=1.2325\times10^{-7}$. A calculation at cap $56$,
$\Delta t=10$, and tolerance $10^{-13}$ ($487{,}635$ states) takes $37.4$
seconds ($66.1$ seconds without the intermediate two-voxel level, see
Supplementary Section~S4.4) and matches
$\mathbb P(k(T)=20)$ to within $0.137\%$ (refinements in Supplementary
Table~S1).

In comparison, $10^6$ independent SSA trajectories~\cite{gillespie1977}
produce three endpoints with $k\ge16$ and none at $k=20$, while
$6\times10^7$ trajectories ($673.5$ seconds) produce $96$ and four
endpoints, respectively. Using the relative standard error
$\sqrt{(1-p)/(mp)}$ for $m$ trajectories, the refined FSP probabilities
predict $9.7\%$ relative standard error for $\mathbb P(k(T)\ge16)$ and
$50\%$ at $k=20$. Reaching $10\%$ relative standard error at $k=20$ with
independent SSA paths would require about $8.1\times10^8$ trajectories,
whereas specialized rare-event sampling methods can reduce variance when
targeted to a reaction coordinate~\cite{allen2005}.

\FloatBarrier
\section{Concluding Remarks}\label{sec:discussion}\label{sec:conclusion}
We have developed a multigrid method for the finite state projection of
the RDME by combining fine-grid relaxation with coarse corrections on
aggregated voxel grids. Summing molecular counts over voxel aggregates
and distributing coarse corrections with multinomial weights preserves
the RDME generator structure and absorbing losses on every level down to
a single-voxel CME. For reversible diffusion, the intergrid transfers
average the error relative to equilibrium across each coarse count class,
and the omitted variance is bounded by the internal aggregate graph
energy with a constant independent of population.

The numerical experiments demonstrate that retaining intermediate spatial
levels reduces iteration counts and runtime relative to direct CME
correction, that W- and K-cycles prevent convergence degradation on deep
one-dimensional hierarchies, and that block Gauss--Seidel relaxation
accelerates convergence on strongly coupled nonlinear populations,
reaching $7.16$ million states and resolving switching probabilities of
order $10^{-7}$. Establishing uniform convergence bounds for the full
reacting cycle and combining voxel aggregation with low-rank state
compression remain natural directions for future work.

\appendix
\section{Proofs of Approximation Properties}\label{app:proofs}
This appendix gives the proofs of Theorem~\ref{thm:reaction},
Lemma~\ref{lem:averaging}, Proposition~\ref{prop:aggregate}, and
Proposition~\ref{prop:residual} stated in Section~\ref{sec:analysis}.

\begin{proof}[Proof of Theorem~\ref{thm:reaction}]
Let $\mathbb E_P[\cdot\mid {X^c}]$ denote expectation under the
prolongation weights \eqref{eq:P} conditioned on a coarse state
${X^c}=X^{c,(m)}$. Under \eqref{eq:P}, the fine count $X_{v,s}$ in voxel
$v\in C_k$ has a binomial marginal distribution with $N={X^c}_{k,s}$
trials and success probability $\theta=\omega_v/\Omega_k$. For
$a=\alpha_{r,s}\le N$, the identity
$\binom{n}{a}\binom{N}{n}=\binom{N}{a}\binom{N-a}{n-a}$ gives the
factorial moment
\[
 \begin{aligned}
 \mathbb E_P\!\left[\binom{X_{v,s}}{\alpha_{r,s}}\,\middle|\,{X^c}\right]
 &=\sum_{n=a}^{N}\binom{n}{a}\binom{N}{n}\theta^n(1-\theta)^{N-n}\\
 &=\binom{N}{a}\theta^a\sum_{n=a}^{N}\binom{N-a}{n-a}\theta^{n-a}(1-\theta)^{N-n}
 =\binom{{X^c}_{k,s}}{\alpha_{r,s}}
       \left(\frac{\omega_v}{\Omega_k}\right)^{\alpha_{r,s}},
 \end{aligned}
\]
which also holds when $N<a$ since both sides vanish. Because the weights
\eqref{eq:P} factor independently across species and
$q_r=\sum_s\alpha_{r,s}$, summing \eqref{eq:voxel-reaction} over
$v\in C_k$ yields
\[
 \begin{aligned}
 \sum_{v\in C_k}\mathbb E_P[a_{r,v}(X)\mid {X^c}]
 &=\sum_{v\in C_k}c_r\omega_v^{1-q_r}
   \prod_s\left[\binom{{X^c}_{k,s}}{\alpha_{r,s}}
   \left(\frac{\omega_v}{\Omega_k}\right)^{\alpha_{r,s}}\right]\\
 &=c_r\Omega_k^{-q_r}\left(\sum_{v\in C_k}\omega_v\right)
   \prod_s\binom{{X^c}_{k,s}}{\alpha_{r,s}}
 =a^c_{r,k}({X^c}).
 \end{aligned}
\]
Every occurrence of reaction $r$ at any fine voxel $v\in C_k$ produces
the same coarse state change $\beta_r-\alpha_r$ in row $k$ of ${X^c}$.
Because $\mathcal X$ is a union of complete population blocks, a fine
destination state belongs to $\mathcal X$ if and only if its coarse
destination belongs to $\mathcal X_c$. An omitted destination state
therefore leaves the averaged loss $-a^c_{r,k}({X^c})$ on the diagonal of
$\mathbf A_c$ without a matching off-diagonal entry.
\end{proof}

\begin{proof}[Proof of Lemma~\ref{lem:averaging}]
Consider a coarse state $X^c=X^{c,(j)}$ at index $j$ with species totals
$k_s=\sum_h X^c_{h,s}$. Let $m(k)>0$ be the positive weight assigned to
population block $k=(k_s)$ and $\Omega=\sum_v\omega_v$. Regrouping the
equilibrium weights \eqref{eq:equilibrium} by aggregate and comparing
with \eqref{eq:P} gives, for every fine state $X^{(i)}$ with $R_{ji}=1$,
\[
 \pi_i=m(k)\prod_s\frac{k_s!}{\Omega^{k_s}}
                 \prod_{h,s}\frac{\Omega_h^{X^c_{h,s}}}{X^c_{h,s}!}
                 \,P_{ij}.
\]
The positive factor preceding $P_{ij}$ depends only on the coarse state
$X^c$. Because $\mathcal X$ retains complete population blocks, the
multinomial theorem implies $\sum_{i:\,R_{ji}=1}P_{ij}=1$. Summing the
factorization of $\pi_i$ over $\{i:R_{ji}=1\}$ therefore identifies the
prefactor as $(R\pi)_j$, and dividing by $(R\pi)_j$ establishes
\eqref{eq:conditional-transfer} with $m(k)$ canceling from the ratio.

For each fine index $i$, let $j$ be its unique coarse index with
$R_{ji}=1$. By \eqref{eq:prolongation-action} and
\eqref{eq:conditional-transfer}, the prolonged restricted error has
entries
\[
 (PRe)_i=P_{ij}(Re)_j=\pi_i\frac{(Re)_j}{(R\pi)_j}.
\]
Substituting $e_i=\pi_i f_i$ and partitioning the sum over fine states by
their coarse index $j$ gives
\[
 \|e-PRe\|_{\pi^{-1}}^2
 =\sum_i\frac{[e_i-(PRe)_i]^2}{\pi_i}
 =\sum_j\ \sum_{i:\,R_{ji}=1}\pi_i
       \left[f_i-\frac{(Re)_j}{(R\pi)_j}\right]^2,
\]
which proves \eqref{eq:variance}.
\end{proof}

\begin{proof}[Proof of Proposition~\ref{prop:aggregate}]
Fix a coarse state $X^c=X^{c,(j)}$ at index $j$ with fine equivalence
class $\mathcal F_j=\{i:R_{ji}=1\}$ and conditional weights
$\mu_j(i)=P_{ij}=\pi_i/(R\pi)_j$. If $|\mathcal F_j|=1$, both its
conditional variance and internal Dirichlet energy vanish. Otherwise,
temporarily label the ${X^c}_{h,s}$ molecules of each species $s$ in each
aggregate $C_h$ and let each labelled molecule diffuse independently on
the connected subgraph $C_h$ with reversible single-molecule generator
$Q_{h,s}$ and stationary distribution $\omega_v/\Omega_h$ ($v\in C_h$).
On the product space $\widetilde{\mathcal F}_j=\prod_{h,s}C_h^{{X^c}_{h,s}}$
of labelled placements $\bm v=(v_{h,s,p})$, the joint stationary measure
and generator are
\[
 \tilde\mu_j(\bm v)=\prod_{h,s}\prod_{p=1}^{{X^c}_{h,s}}
 \frac{\omega_{v_{h,s,p}}}{\Omega_h},\qquad
 \widetilde Q_j=\sum_{h,s}\sum_{p=1}^{{X^c}_{h,s}}
 I\otimes\cdots\otimes Q_{h,s}^{(p)}\otimes\cdots\otimes I.
\]
Because the eigenvalues of $-\widetilde Q_j$ are sums of the nonnegative
eigenvalues of $-Q_{h,s}^{(p)}$, the smallest positive eigenvalue of
$-\widetilde Q_j$ is at least $\lambda_*$. The spectral expansion on
$L^2(\widetilde{\mathcal F}_j,\tilde\mu_j)$ therefore yields the
Poincar\'e inequality
\[
 \sum_{\bm v\in\widetilde{\mathcal F}_j}\tilde\mu_j(\bm v)
 \left[\tilde f(\bm v)-\sum_{\bm w}\tilde\mu_j(\bm w)\tilde f(\bm w)\right]^2
 \le\frac{1}{2\lambda_*}\sum_{\bm v\ne\bm w}
 \tilde\mu_j(\bm v)(\widetilde Q_j)_{\bm w,\bm v}
 \left[\tilde f(\bm v)-\tilde f(\bm w)\right]^2.
\]
Evaluating this inequality on functions $\tilde f(\bm v)=f_i$ that depend
only on the unlabelled voxel counts $X^{(i)}$ sums $\tilde\mu_j(\bm v)$
over the $\prod_{h,s}({X^c}_{h,s}!/\prod_{v\in C_h}X^{(i)}_{v,s}!)$
permutations in each count state $i\in\mathcal F_j$ to recover the
multinomial weight $\mu_j(i)=P_{ij}$ and the count-space jump rates
$d_{vw}^{(s)}X^{(i)}_{v,s}=(\mathbf A_{\mathrm d})_{mi}$. Consequently,
\[
 \sum_{i\in\mathcal F_j}\mu_j(i)
 \left[f_i-\frac{(Re)_j}{(R\pi)_j}\right]^2
 \le\frac{1}{2\lambda_*}\sum_{\substack{i,m\in\mathcal F_j\\ i\ne m}}
  \mu_j(i)(\mathbf A_{\mathrm d})_{mi}(f_i-f_m)^2.
\]
Multiplying by $(R\pi)_j$ and summing over all coarse classes $j$ yields
$\|e-PRe\|_{\pi^{-1}}^2\le\lambda_*^{-1}\mathcal E_{\mathrm{in}}(f)$ by
\eqref{eq:variance}.
\end{proof}

\begin{proof}[Proof of Proposition~\ref{prop:residual}]
For $\alpha=\max_i(-\mathbf A_{ii})\ge0$, the shifted matrix
$\mathbf A+\alpha I$ has nonnegative entries, so
$e^{t\mathbf A}=e^{-\alpha t}e^{t(\mathbf A+\alpha I)}$ is entrywise
nonnegative for all $t\ge0$. Since $\one^\top\mathbf A\le0$,
differentiating $\one^\top e^{t\mathbf A}$ shows that the column sums of
$e^{t\mathbf A}$ do not exceed one, giving $\|e^{t\mathbf A}\|_1\le1$.
For $\sigma>0$, the Laplace integral for the resolvent gives
\[
 \mathbf A_\sigma^{-1}=\int_0^\infty e^{-\sigma t}e^{t\mathbf A}\,dt,
 \qquad
 \|\mathbf A_\sigma^{-1}\|_1
 \le\int_0^\infty e^{-\sigma t}\|e^{t\mathbf A}\|_1\,dt
 \le\sigma^{-1}.
\]
Applying this operator norm bound to
$x-\mathbf A_\sigma^{-1}b=-\mathbf A_\sigma^{-1}(b-\mathbf A_\sigma x)$
establishes \eqref{eq:residual}.
\end{proof}

\section*{Code Availability}
The solver and reproduction scripts are available at
\url{https://github.com/AdityaDendukuri/markovkit/tree/main/rdme}.

\Needspace{8\baselineskip}

\end{document}

\maketitle
Section~\ref{sec:worked-example} walks through an explicit algebraic
example of the state space and transfer operators on two voxels,
followed by a coarse correction step. Section~\ref{sec:multigrid-analysis}
derives the orthogonalized K-cycle coefficients, the two-grid energy
optimality under detailed balance, and the operation count per cycle.
Section~\ref{sec:numerical-validation} presents independent
uniformization checks and the adaptive population enlargement algorithm,
together with the discretization refinement study for rare switching.
Section~\ref{sec:implementation} examines matrix storage, transfer
sparsity patterns, and exact reduction under reflection symmetry, while
Section~\ref{sec:spatial-front} illustrates spatial front propagation in
a bistable Schl\"ogl chain using exact trajectory simulations. We retain
all notation and solver conventions of the main paper, and any section
or equation number without an S prefix refers to the main paper.

\section{A Worked Example}\label{sec:worked-example}
A linear immigration and degradation network illustrates how the
generator and transfer operators are constructed on a domain of two
voxels, together with the resulting Galerkin coarse system. As in the
main paper, columns of the generator correspond to source states.

\subsection{State Space and Generator}
Consider a single chemical species $\mathrm S$ in a compartment of unit
volume ($\omega=1$), subject to immigration $\varnothing\to \mathrm{S}$
with propensity $1$ and degradation $\mathrm{S}\to\varnothing$ with
propensity $X$. In this single-voxel CME, the states are the integer copy
numbers $X\in\{0,1,2,\ldots\}$, and the probability vector has one
component for each count. An immigration event transfers probability flux
from component $X$ to component $X+1$.

Partitioning the unit domain into two voxels of equal volume
$\omega_1=\omega_2=1/2$ yields spatial configurations $X=(X_1,X_2)$ with
joint probabilities $p(X_1,X_2,t)$. In each voxel $v\in\{1,2\}$,
immigration occurs at rate $1/2$ and degradation occurs at rate $X_v$.
Let $d>0$ denote the diffusive jump rate per molecule between the two
voxels. An immigration event in voxel $1$ transitions state $(1,0)$ to
$(2,0)$ at rate $1/2$, whereas a diffusive jump transitions $(2,0)$ to
$(1,1)$ at rate $2d$.

Truncating the state space by total population $X_1+X_2\le {N_{\max}}$
with cap ${N_{\max}}=2$ retains the six configurations
\begin{equation}\label{eq:running-states}
 (0,0),\quad(1,0),\quad(0,1),\quad
 (2,0),\quad(1,1),\quad(0,2).
\end{equation}
Ordering the probability vector according to \eqref{eq:running-states}
groups states by total population $k=0,1,2$. The configuration $(2,2)$
is excluded because the FSP cap bounds the total copy number $X_1+X_2$
rather than each voxel count separately. Any immigration event occurring
in the $k=2$ block transitions to a state with $k=3$ outside the
retained set, so its outgoing rate $1$ remains on the diagonal as an
absorbing loss. Assembling all reaction and diffusion transitions yields
the $6\times6$ FSP generator
\begin{equation}\label{eq:example-fine-generator}
\mathbf A=\begin{pmatrix}
 -1&1&1&0&0&0\\
 1/2&-2-d&d&2&1&0\\
 1/2&d&-2-d&0&1&2\\
 0&1/2&0&-3-2d&d&0\\
 0&1/2&1/2&2d&-3-2d&2d\\
 0&0&1/2&0&d&-3-2d
\end{pmatrix}.
\end{equation}
The first three columns correspond to states with $X_1+X_2<2$ and have
column sums equal to zero, whereas the last three columns correspond to
$X_1+X_2=2$ and have column sums equal to $-1$ due to the absorbed
immigration flux. Increasing the cap from ${N_{\max}}=2$ to
${N_{\max}}=3$ appends the four states $(3,0),(2,1),(1,2),(0,3)$,
enlarging the probability vector from six to ten components while keeping
two voxel coordinates per state.

\subsection{Restriction and Prolongation}
Aggregating the two voxels into a single coarse compartment of volume
$\Omega_1=\omega_1+\omega_2=1$ gives the coarse count
${X^c}=X_1+X_2\in\{0,1,2\}$. The restriction operator $R$ sums entries
over fine configurations with the same total ${X^c}$, while the
prolongation operator $P$ distributes each molecule independently into
voxel $1$ or voxel $2$ with equal probability $\omega_v/\Omega_1=1/2$.
In the state ordering \eqref{eq:running-states}, these operators take
the block-diagonal form
\begin{equation}\label{eq:running-prolongation}
 R=\begin{pmatrix}
 1&0&0&0&0&0\\
 0&1&1&0&0&0\\
 0&0&0&1&1&1
 \end{pmatrix},\qquad
 P=\begin{pmatrix}
 1&0&0\\
 0&1/2&0\\
 0&1/2&0\\
 0&0&1/4\\
 0&0&1/2\\
 0&0&1/4
 \end{pmatrix}.
\end{equation}
In the solver implementation, the nonzeros of $R$ and $P$ are stored in
the two length-$6$ arrays
\[
 \texttt{parent}=(0,1,1,2,2,2),\qquad
 \texttt{weight}=(1,1/2,1/2,1/4,1/2,1/4),
\]
using zero-based coarse indices. For an example fine probability vector,
restriction sums within each population block according to
\begin{equation}\label{eq:running-probabilities}
 \begin{array}{c|c|c|c}
 {X^c}&\text{fine states}&\text{fine probabilities}&p_c({X^c})\\ \hline
 0&(0,0)&0.10&0.10\\
 1&(1,0),(0,1)&0.10,\ 0.10&0.20\\
 2&(2,0),(1,1),(0,2)&0.14,\ 0.35,\ 0.21&0.70
 \end{array}
\end{equation}
Applying the projection $PR$ to this probability vector gives
\[
 \begin{aligned}
 \boldsymbol p&=(0.10,0.10,0.10,0.14,0.35,0.21)^\top,\\
 PR\boldsymbol p&=(0.10,0.10,0.10,0.175,0.35,0.175)^\top.
 \end{aligned}
\]
The projected vector $PR\boldsymbol p$ preserves the coarse marginal
$R\boldsymbol p$ while replacing the asymmetric entries on the $k=2$
block by the binomial values $0.70\times(1/4,1/2,1/4)^\top$. Although
$RP=I$ ensures that restricting a prolonged vector recovers the coarse
vector, $PR$ has rank three and cannot recover the fine components in
the null space of $R$.

This two-voxel system can itself serve as an intermediate level in a
deeper hierarchy. Starting from a chain of four voxels at cap
${N_{\max}}=2$ ($15$ states), pairing voxels $\{1,2\}$ and $\{3,4\}$
maps a fine state such as $(1,0,0,1)$ to the two-voxel state $(1,1)$ and
then to the single-voxel CME state $2$, reducing the state-space
dimension through $15\to6\to3$ while retaining all populations up to
cap two on every level.

\subsection{The Coarse Generator}
Forming the Galerkin projection of \eqref{eq:example-fine-generator} with
the transfer matrices \eqref{eq:running-prolongation} produces the
$3\times3$ coarse generator
\begin{equation}\label{eq:running-coarse-generator}
 \mathbf A_c=R\mathbf AP=\begin{pmatrix}
 -1&1&0\\
 1&-2&2\\
 0&1&-3
 \end{pmatrix}.
\end{equation}
All diffusive parameter $d$ terms cancel identically in $R\mathbf AP$
because diffusive jumps remain within the single aggregate $\{1,2\}$. The
coarse immigration rates sum to $1/2+1/2=1$, the coarse degradation
rates sum to $X_1+X_2={X^c}$, and the third column retains its column
sum of $-1$ from the absorbed immigration channel at ${X^c}=2$. Thus
$\mathbf A_c$ is the exact FSP generator of the single-voxel CME at cap
two.

When a two-voxel grid arises by coarsening a four-voxel chain of uniform
spacing $h$ and fine jump rate $D/h^2$, only the boundary voxel inside
each two-voxel aggregate is adjacent to the interface between the two
aggregates. Because a molecule in either aggregate occupies that
boundary voxel with probability $\omega_v/\Omega_k=1/2$, the Galerkin
coarse jump rate \textup{(21)} between the two
aggregates is $d^c=(1/2)(D/h^2)=D/(2h^2)$. By contrast, naive finite-difference
rediscretization at coarse mesh spacing $2h$ with diffusivity $D$ would
assign jump rate $D/(2h)^2=D/(4h^2)$, differing by a factor of two from
the Galerkin rate $R\mathbf AP$.

\subsection{A Coarse Correction}
To trace a single coarse-grid correction step, consider backward Euler
with step size $\Delta t=1$ ($\sigma=1$) starting from the empty state
$b=\boldsymbol p(0)=(1,0,0,0,0,0)^\top$. Initializing the solve with
$x=b$ for $\mathbf A_\sigma=I-\mathbf A$ gives the fine residual and its
restriction
\[
 r=b-\mathbf A_\sigma x=(-1,1/2,1/2,0,0,0)^\top,\qquad
 Rr=(-1,1,0)^\top.
\]
Solving the shifted coarse equation
$(I-\mathbf A_c)e_c=Rr$ yields
\[
 \begin{pmatrix}
 2&-1&0\\
 -1&3&-2\\
 0&-1&4
 \end{pmatrix}e_c
 =\begin{pmatrix}-1\\1\\0\end{pmatrix},\qquad
 e_c=\begin{pmatrix}-3/8\\1/4\\1/16\end{pmatrix}.
\]
Prolonging $e_c$ by $P$ and adding the correction to $x$ produces the
updated fine iterate
\[
 x^+=x+Pe_c=(5/8,1/8,1/8,1/64,1/32,1/64)^\top.
\]
Direct multiplication verifies that $(I-\mathbf A)x^+=b$ for every
$d>0$, with retained probability mass $\one^\top x^+=15/16$. Because the
initial state and linear reaction propensities in this spatially
symmetric model preserve the binomial distribution across the two
voxels, the exact solution lies entirely in the range of $P$ and is
recovered in a single coarse correction without relaxation.

\subsection{Nonlinear Reactions and Coarse Approximation}
When nonlinear reactions are present, the fine solution no longer lies
in the range of $P$. Suppose we add the bimolecular birth reaction
$2\mathrm{S}\to3\mathrm{S}$ with rate constant $c>0$ in each voxel of
volume $1/2$. Its combined propensity across the two voxels is
\[
 2c\left[\binom{X_1}{2}+\binom{X_2}{2}\right],
\]
which evaluates to $2c$, $0$, and $2c$ on the three states
$(2,0)$, $(1,1)$, and $(0,2)$ of the $k=2$ block. Averaging these three
values with the binomial prolongation weights $(1/4,1/2,1/4)^\top$ gives
the coarse propensity $c\binom{{X^c}}{2}=c$ at ${X^c}=2$, in agreement
with Theorem~4.1. The true outgoing flux from the
$k=2$ block, however, depends on how much probability resides in $(1,1)$
versus $(2,0)$ and $(0,2)$, so the coarse counts do not evolve
autonomously and the solver must iterate on the fine system.

On the $k=2$ block, an error subvector proportional to the binomial
weights $(1,2,1)^\top$ lies in the range of $P$ and is reproduced
exactly by $PR$, whereas an oscillatory error subvector $(1,-2,1)^\top$
lies in the null space of $R$ and is omitted by $PR$.
Proposition~4.3 bounds this omitted component in
terms of the Dirichlet energy of diffusive jumps between the two voxels,
while relaxation sweeps on $\mathbf A_\sigma$ damp $(1,-2,1)^\top$
directly on the fine grid.

\section{Multigrid Analysis}\label{sec:multigrid-analysis}
Among the analytical results below, only the orthogonality argument in
Section~\ref{sec:projection} assumes detailed balance for the full
reacting FSP generator.

\subsection{K-Cycle Coefficients}\label{app:kcycle}
The K-cycle chooses coefficients for recursive corrections by minimizing
the coarse Euclidean residual norm. Let $b_c$ be the restricted residual
from Section~3.6 and let
$\mathbf A_{\sigma,c}=\mathbf A_{\sigma,l +1}$. A zero right-hand side
returns zero, and a direction with zero image under $\mathbf A_{\sigma,c}$
leaves the correction unchanged, so we evaluate the following weights
only when their denominators are nonzero. The first recursive direction
and the residual after optimal scaling are
\begin{equation}\label{eq:kfirst}
 v_1=\Cycle( l +1,b_c),\qquad
 \alpha_1=\frac{(\mathbf A_{\sigma,c}v_1)^\top b_c}{\|\mathbf A_{\sigma,c}v_1\|_2^2},
 \qquad r_1=b_c-\alpha_1\mathbf A_{\sigma,c}v_1.
\end{equation}
When $\|r_1\|_2\le\|b_c\|_2/4$, the correction $e_c=\alpha_1v_1$ is
accepted. Otherwise, we compute a second recursive direction from $r_1$
and subtract its component along the first so that their images under
$\mathbf A_{\sigma,c}$ are orthogonal,
\begin{equation}\label{eq:ksecond}
 \begin{aligned}
 v_2&=\Cycle( l +1,r_1),&
 \beta&=\frac{(\mathbf A_{\sigma,c}v_1)^\top\mathbf A_{\sigma,c}v_2}{\|\mathbf A_{\sigma,c}v_1\|_2^2},&
 v_2'&=v_2-\beta v_1,\\
 \alpha_2&=\frac{(\mathbf A_{\sigma,c}v_2')^\top r_1}{\|\mathbf A_{\sigma,c}v_2'\|_2^2},&&
 e_c&=\alpha_1v_1+\alpha_2v_2'.
 \end{aligned}
\end{equation}
The orthogonality relations
$(\mathbf A_{\sigma,c}v_1)^\top\mathbf A_{\sigma,c}v_2'=0$ and
$(\mathbf A_{\sigma,c}v_1)^\top r_1=0$ imply that these coefficients
minimize $\|b_c-\mathbf A_{\sigma,c}e_c\|_2$ over the span of the two
directions. At the bottom CME level, the direct solve replaces this
minimization.

Equations~\eqref{eq:kfirst} and~\eqref{eq:ksecond} perform at most two
steps of preconditioned generalized conjugate residual
minimization~\cite{notay2008}, whose dependence on $b_c$ makes the
cycle nonlinear. We therefore apply the K-cycle directly as an iterative
solver and use the linear W-cycle when right-preconditioning
GMRES~\cite{saad1986}.

\Needspace{8\baselineskip}
\subsection{Coarse Correction under Detailed Balance}\label{sec:projection}
The coarse representation $PRe$ and the Galerkin coarse correction both
lie in the range of $P$, but their coefficients are determined
differently. Whereas $PRe$ uses restriction alone, the coarse correction
solves the Galerkin system for the restricted residual. For
$\mathbf A_\sigma =\sigma I-\mathbf A$ with $\sigma>0$, an exact coarse
solve maps an error $e$ to $Ce$, where
\[
 C=I-P\mathbf A_{\sigma,c}^{-1}R\mathbf A_\sigma ,\qquad\mathbf A_{\sigma,c}=R\mathbf A_\sigma P.
\]
The identities $C^2=C$, $CP=0$, and $R\mathbf A_\sigma C=0$ show that an
exact coarse correction removes any error component in the range of $P$
and leaves zero restricted residual. Under detailed balance, the standard
Galerkin argument~\cite{xu1992} shows that this correction minimizes a
weighted operator energy norm.

Let $\pi$ be positive weights on the retained fine states, with
$\Pi=\operatorname{diag}(\pi)$ and $\Pi_c=\operatorname{diag}(R\pi)$, and
retain $f_i=e_i/\pi_i$. For example, linear immigration at rate
$\eta\omega_v$ and degradation at rate $\mu X_v$ ($\eta,\mu>0$) under
reversible diffusion admit independent Poisson weights with means
$\eta\omega_v/\mu$~\cite{anderson2010}, whose conditional distribution on
the aggregate counts equals the prolongation weights $P$.

Assume that $\mathbf A_{ji}\pi_i=\mathbf A_{ij}\pi_j$ holds for the
complete retained generator and that $\pi_i/(R\pi)_j=P_{ij}$ whenever
$R_{ji}=1$. The weighted operator $H=\Pi^{-1}\mathbf A_\sigma$ is then
symmetric and satisfies
\begin{equation}\label{eq:weighted-galerkin}
 \Pi_c^{-1}R=P^\top \Pi^{-1},\qquad
 \Pi_c^{-1}\mathbf A_{\sigma,c}=P^\top HP.
\end{equation}
Since the absorbing loss rate from state $X^{(i)}$ to omitted states is
$-\sum_j\mathbf A_{ji}\ge0$, the quadratic form of $H$ in terms of the
relative error $f$ is
\[
 \begin{aligned}
 e^\top He
 &=\sigma\sum_i\pi_i f_i^2
   +\frac12\sum_{i\ne j}\pi_i\mathbf A_{ji}
                         (f_j-f_i)^2\\
 &\quad+\sum_i\pi_i\left(-\sum_j\mathbf A_{ji}\right)f_i^2>0\quad(e\ne0).
 \end{aligned}
\]
The shift, transitions between retained states, and absorbing losses each
contribute nonnegative terms, so $\|e\|_H^2=e^\top He$ is an energy
norm on $\mathcal X$. If $\mathbf A_{\mathrm{d,in}}$ retains only
diffusive jumps inside aggregates, Proposition~4.3
gives
\[
 \|e-PRe\|_{\pi^{-1}}^2
 \le\lambda_*^{-1}e^\top \Pi^{-1}(-\mathbf A_{\mathrm{d,in}})e
 \le\lambda_*^{-1}e^\top He.
\]
By \eqref{eq:weighted-galerkin}, the corrected error $Ce$ is orthogonal
in the $H$-norm to the range of $P$, giving
\[
 P^\top HC=\Pi_c^{-1}R\mathbf A_\sigma (I-P\mathbf A_{\sigma,c}^{-1}R\mathbf A_\sigma )=0,
 \qquad
 \|e\|_H^2=\|Ce\|_H^2+\|(I-C)e\|_H^2,
\]
and hence
\[
 \|Ce\|_H=\min_{z_c}\|e-Pz_c\|_H.
\]
An exact coarse solve therefore selects the prolonged correction that
minimizes the remaining error in the $H$-norm.

The point Gauss--Seidel smoother of Section~3.5
also reduces this norm. For the error $e$ before updating component $i$
and $e^+$ afterward, $(He)_i/H_{ii}=(\mathbf A_\sigma e)_i/(\mathbf A_\sigma)_{ii}$
gives
\[
 e_i^+=e_i-\frac{(He)_i}{H_{ii}},\qquad e_j^+=e_j\quad(j\ne i),
 \qquad
 \|e^+\|_H^2=\|e\|_H^2-\frac{(He)_i^2}{H_{ii}}.
\]
Equality over a full sweep requires every update to vanish, implying
$He=0$ and $e=0$. Compactness of the unit energy sphere then implies
strict two-grid contraction on each finite system, with a factor that may
depend on system size.

In a multilevel cycle, let $F_ l $ and $B_ l $ denote the error
propagators of forward and reverse relaxation on level $l$, and let
$E_ l $ denote the error propagator of the complete cycle. The direct
bottom solve gives $E_{N_{\mathrm l}}=0$, while the V- and W-cycles
satisfy
\[
 E_ l =B_ l \left[I-P_ l (I-E_{ l +1}^{\gamma})
         \mathbf A_{\sigma,l +1}^{-1}R_ l \mathbf A_{\sigma,l}\right]F_ l ,
 \qquad\gamma=1\text{ or }2,
\]
with $\gamma=1$ for V and $\gamma=2$ for W. Because the K-cycle weights
depend on the residual, K has no single linear error propagator, and none
of these identities alone establishes a uniform convergence rate.

The nonlinear Schl\"ogl experiments do not assume detailed balance for
the complete generator, so this $H$-norm argument does not apply to them.
Their diffusion part still has the graph energy in
\textup{(10)}, and the transfer operators, coarse
rates, and residual bound of Proposition~4.4 hold
without reversibility of the full system.

\subsection{Matrix Size and Work per Cycle}\label{app:work}
The work estimate in Section~3.7 follows from the
number of nonzeros on each level and the number of recursive visits.
Consider a birth--death system for one species on ${N_{\mathrm v}}$
voxels at cap ${N_{\max}}\ge1$, and let $\mathcal E$ be the edge set of
the voxel graph. Each edge permits diffusion in both directions, so when
all permitted event rates are positive the shifted matrix has
\[
 \nnz(\mathbf A_\sigma )=\binom{{N_{\mathrm v}}+{N_{\max}}}{{N_{\max}}}
                  +(2{N_{\mathrm v}}+2|\mathcal E|)\binom{{N_{\mathrm v}}+{N_{\max}}-1}{{N_{\max}}-1}.
\]
For each voxel, $\binom{{N_{\mathrm v}}+{N_{\max}}-1}{{N_{\max}}-1}$
states contain at least one molecule there and the same number lie below
the cap and permit a birth. On a $d$-dimensional Cartesian grid with
$2|\mathcal E|\le2d{N_{\mathrm v}}$, this yields
\[
 \nnz(\mathbf A_\sigma )\le[1+(2+2d){N_{\max}}]\binom{{N_{\mathrm v}}+{N_{\max}}}{{N_{\max}}}.
\]
For fixed ${N_{\max}}$, the state count is asymptotic to
${N_{\mathrm v}}^{N_{\max}}/{N_{\max}}!$, so coarsening by $2^d$ voxels
reduces the leading term by $2^{-d{N_{\max}} l }$ on level $ l $. With at
most $2^ l $ visits per W-cycle, the total work is bounded by a geometric
sum of ratio $2^{1-d{N_{\max}}}$, which is $O(N)$ independently of the
number of levels when $d{N_{\max}}>1$ and $O(N\log N)$ when
$d={N_{\max}}=1$. The V- and K-cycles have the same asymptotic work bound
per cycle, excluding setup and the number of cycles required for
convergence.

\section{Numerical Validation}\label{sec:numerical-validation}
These checks assess numerical accuracy separately from the cost
comparisons in Section~\ref{sec:implementation}.

\subsection{Uniformization}\label{app:validation}
On smaller state spaces we check the time integration independently by
uniformization~\cite{sidje1999}, which evaluates the matrix exponential as
a Poisson-weighted series of powers of a shifted transition matrix
without solving linear systems. For the larger four-voxel systems, we
instead use the population and time refinements in
Section~5.4.

For the Schl\"ogl transient on two voxels at cap $40$, uniformization
with subintervals corresponding to Poisson means $4$ and $8$ produces
reference probability vectors agreeing to $6.45\times10^{-13}$ in
$\ell_1$. Relative to this reference at $T=100$, the Crank--Nicolson
integrator with backward Euler startup has $\ell_1$ errors
$2.38\times10^{-3}$, $4.13\times10^{-4}$, $1.40\times10^{-4}$, and
$2.26\times10^{-5}$ at $\Delta t=10$, $5$, $2.5$, and $1.25$,
respectively.

\subsection{Adaptive Population Selection}\label{app:adaptive}
The adaptive FSP algorithm selects the population cap from a cumulative
$\ell_1$ error budget while retaining complete population blocks on every
level. At each trial step, we apply backward Euler to the current FSP
system and bound the algebraic solve error using
Proposition~4.4. Because the exact backward Euler
resolvent $(I-\Delta t\mathbf A)^{-1}$ is contractive in the $\ell_1$
norm, the sum of residual bounds over accepted steps bounds the
accumulated algebraic error of the linear solves.

A trial step is accepted when its computed missing mass $1-\one^\top x^m$
plus twice this accumulated sum of solve bounds (including the current
trial bound) does not exceed the total error budget multiplied by the
elapsed time fraction $t_m/T$. One copy of the accumulated solve bound
accounts for the difference between the computed missing mass and the
exact FSP truncation loss, and the second copy bounds the algebraic error
in the returned vector $x^m$. Rejected trial steps do not contribute to
the accumulated sum.

Whenever a trial step violates this acceptance bound, the driver
increases the population cap from ${N_{\max}}$ to
$\max\{{N_{\max}}+4,\lceil1.25({N_{\max}}+1)\rceil\}$ across all levels
of the spatial hierarchy and repeats the step from the saved distribution
$x^{m-1}$, padding the newly added population blocks with zeros without
renormalizing previously absorbed mass. All existing diagonal blocks
$(\mathbf A_{\sigma,l})_{kk}$ and their ILU(0) factors are retained
without recomputation, along with the existing transfer entries. This
procedure guarantees that the accepted discrete trajectory satisfies
$\|\boldsymbol p_*^m-x^m\|_1\le\varepsilon_{\mathrm{FSP}}(t_m/T)$ at
every step, while temporal discretization error relative to the
continuous-time RDME is controlled through $\Delta t$.

\subsection{Refinement of the Rare-Switching Calculation}
Table~\ref{tab:rare-refinement} reports the cap and step-size
refinements for the rare-switching problem of
Section~5.5 on four voxels at $T=500$. Every
linear solve converges to the prescribed tolerance ($10^{-13}$ at cap
$56$ and $10^{-14}$ at caps $64$ and $72$), and all final probability
entries are nonnegative without clipping or renormalization. These
sequential timings include solver setup and file output on the benchmark
hardware of Section~5 using the unreduced state
representation, and are separate from the paired implementation
comparisons in Section~\ref{sec:implementation-results}.

\begin{table}[!htbp]
\centering
\caption{Discretization refinement for the rare-switching RDME on four voxels
at $T=500$. All runs use the $4\to2\to1$ W-cycle with one forward and
one reverse ILU(0) block sweep per level, with solve tolerance
$10^{-13}$ at cap 56 and $10^{-14}$ at caps 64 and 72.}
\label{tab:rare-refinement}
\begingroup\small\setlength{\tabcolsep}{4pt}
\begin{tabular}{rrrrrr}
\toprule
cap & $\Delta t$ & states & $\mathbb P(k\ge16)$ & $\mathbb P(k=20)$ & time (s)\\
\midrule
56 & 10 & 487,635 & $1.7719\times10^{-6}$ & $1.2308\times10^{-7}$ & 37.4\\
64 & 5 & 814,385 & $1.7746\times10^{-6}$ & $1.2325\times10^{-7}$ & 134.4\\
64 & 2.5 & 814,385 & $1.7746\times10^{-6}$ & $1.2325\times10^{-7}$ & 231.3\\
72 & 5 & 1,282,975 & $1.7746\times10^{-6}$ & $1.2325\times10^{-7}$ & 209.3\\
\bottomrule
\end{tabular}\endgroup
\end{table}
\FloatBarrier

\section{Implementation and Computational Cost}\label{sec:implementation}
\subsection{Matrix Storage and Factor Reuse}\label{sec:implementation-details}
To minimize memory overhead and setup time, the solver exploits the
bipartite graph structure of Cartesian diffusion blocks during ILU(0)
factorization. On a bipartite voxel graph (such as any Cartesian mesh under
a red--black checkerboard coloring), every diffusive jump transfers a
molecule between a red voxel and a black voxel, changing the parity of
the total molecule count on the red voxels by $\pm1$. Consequently, the
diffusion transition graph within each population block is itself
bipartite and contains no odd cycles, and in particular no triangles
($3$-cycles). During an ILU(0) factorization of a diagonal block
$(\mathbf A_{\sigma,l})_{kk}$, an off-diagonal entry $(i,j)$ is modified
by elimination of a prior pivot $m$ only if $(i,m)$, $(m,j)$, and
$(i,j)$ are all nonzero, which would form a triangle in the diffusion
graph. Because no such triangles exist, all off-diagonal entries of the
ILU(0) factors coincide identically with the off-diagonal entries of
$(\mathbf A_{\sigma,l})_{kk}$, and only the modified diagonal pivots
need to be computed and stored. Forward and backward triangular solves
in \textup{(26)} therefore read the off-diagonal entries
directly from $(\mathbf A_{\sigma,l})_{kk}$, avoiding duplicate sparse
matrix storage while producing the exact same ILU(0) update in
floating-point arithmetic. We also fuse the block residual evaluation
with the forward sweep and reuse workspace buffers across time steps.

\subsection{Structure of the Transfer Operators}
To visualize the algebraic structure of the transfer operators in the
Schl\"ogl benchmark on four voxels at cap $40$ (Table~3),
we export the arrays \texttt{parent} and \texttt{weight} from the
$4\to2\to1$ hierarchy ($135{,}751\to861\to41$ states) and plot all
nonzero locations in Supplementary Figure~\ref{fig:transfer-sparsity}.
In the solver's native ordering, states are sorted first by total
population $k=0,\ldots,40$ and then lexicographically by voxel counts.
Because $R_l$ and $P_l$ conserve total population, both matrices are
block diagonal across the $41$ population blocks, and the sparsity
pattern of $R_l$ is the exact transpose of that of $P_l$.

\begin{figure}[!htbp]
\centering
\includegraphics[width=\linewidth]{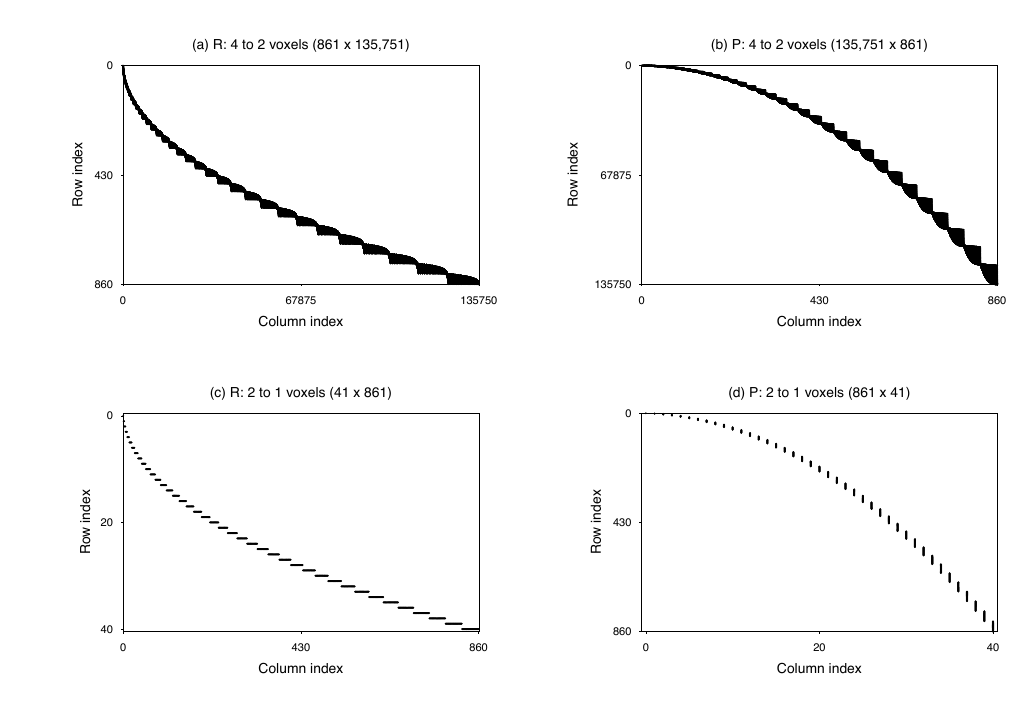}
\caption{Sparsity patterns of the restriction and prolongation operators from
the Schl\"ogl benchmark on four voxels at cap 40. Panels (a,b) display
the $4\to2$ transfers ($135{,}751$ nonzeros each), and panels (c,d)
display the $2\to1$ transfers ($861$ nonzeros each) in the solver's
native ordering. Every nonzero entry is plotted as a black point using
zero-based indices. The upward curvature reflects the cubic growth of
the population block sizes $\binom{k+3}{3}$ on four voxels.}
\label{fig:transfer-sparsity}
\end{figure}

\begin{figure}[!htbp]
\centering
\includegraphics[width=\linewidth]{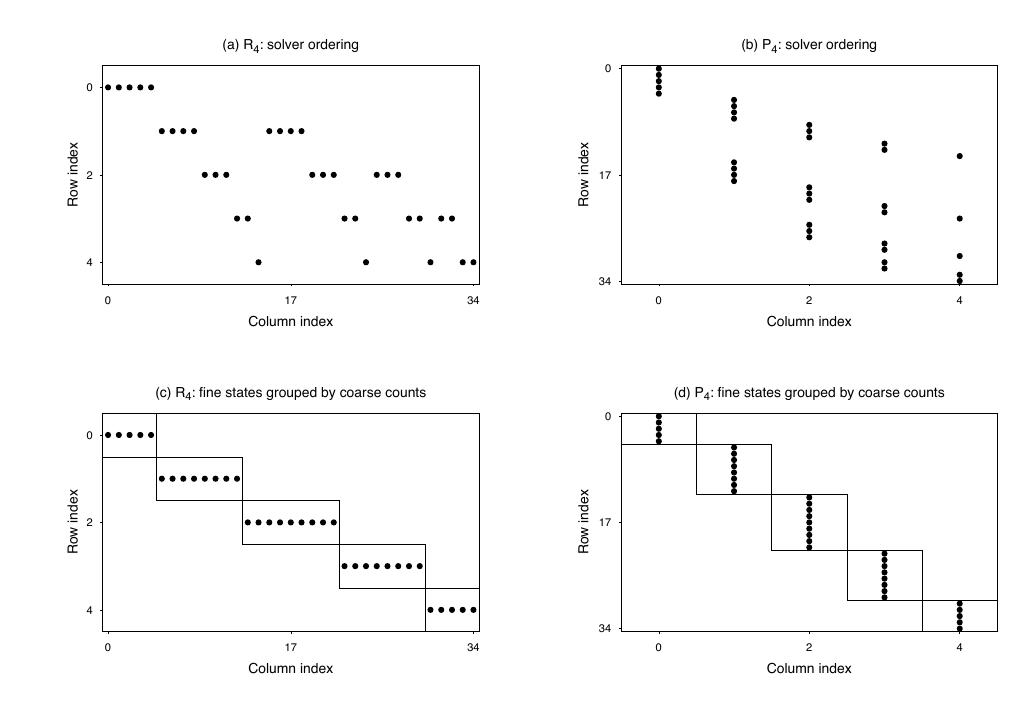}
\caption{Sparsity pattern of the $k=4$ population block ($35$ fine states and
$5$ coarse states) of the $4\to2$ transfer operators. Panels (a,b) use
the solver's native lexicographic ordering, whereas panels (c,d) group
fine states by their coarse counts to display the five disjoint
rectangular blocks of \textup{(15)}. Restriction has
dimension $5\times35$ and prolongation has dimension $35\times5$, with
$35$ nonzeros in each panel.}
\label{fig:transfer-block}
\end{figure}

Supplementary Figure~\ref{fig:transfer-block} zooms into the single
population block $k=4$ of the $4\to2$ transfer operators, which maps
$35$ fine states on four voxels to $5$ coarse states on two voxels. In
the native lexicographic ordering (panels a and b), fine states belonging
to different coarse states $(X^c_1,X^c_2)$ are interleaved. Permuting the
$35$ fine states so that arrangements sharing each of the five coarse
states $(4,0)$, $(3,1)$, $(2,2)$, $(1,3)$, and $(0,4)$ are contiguous
(panels c and d) reveals five disjoint rectangular blocks of sizes
$5,8,9,8,5$ as in \textup{(15)}. Within each
block, restriction sums the fine entries with a row of ones, and
prolongation distributes the coarse entry with the corresponding column
of multinomial weights.

\subsection{Exact Reduction by Reflection Symmetry}\label{sec:reflection}
When the model and initial distribution are invariant under spatial
reflection, the FSP state space can be compressed exactly by grouping
configurations that are related by reflection. On a chain of
$N_{\mathrm v}$ voxels of equal volume with spatially uniform rates,
reversing the voxel index map $v\mapsto N_{\mathrm v}+1-v$ commutes with
the FSP generator $\mathbf A$ (and similarly along each coordinate axis
on a $d$-dimensional Cartesian grid). Consequently, any configuration $X$
and its reflection $\widetilde X$ carry identical probability
$p(X,t)=p(\widetilde X,t)$ at all times $t\ge0$. Storing the combined
probability $p(X,t)+p(\widetilde X,t)$ as a single unknown for each
asymmetric pair $\{X,\widetilde X\}$ and storing symmetric configurations
($X=\widetilde X$) once halves the asymptotic dimension of the system,
while the full probability vector is recovered exactly by dividing each
paired entry by two.

Summing the outgoing transition rates from a representative source state
$X$ into each destination group gives the exact reduced generator with
the same absorbing boundary losses. Because uniform $2^d$ voxel
aggregation commutes with spatial reflection, restriction still sums
reduced entries and prolongation weights each fine group by its summed
multinomial weight. Unlike the bipartite unreduced graph in
Section~\ref{sec:implementation-details}, identifying reflected states
can introduce triangles into the reduced diffusion graph, so the reduced
solver uses a general sparse ILU(0) factorization on each diagonal
population block while keeping the bottom CME on one voxel unchanged.

\subsection{Computational Cost}\label{sec:implementation-results}
We now evaluate the memory and runtime reductions obtained from the
ILU(0) storage of Section~\ref{sec:implementation-details} and the
reflection reduction of Section~\ref{sec:reflection} while holding the
FSP system and multigrid parameters fixed. All runs use the $4\to2\to1$
W-cycle with one forward and one reverse block Gauss--Seidel sweep per
level. Each paired comparison executes two sequential repetitions in
alternating order on the same machine, and all reported elapsed times
include setup and time integration together with exporting the complete
unreduced probability vector.

On the rare-switching integration on four voxels at cap $56$ without
reflection reduction, eliminating duplicate off-diagonal ILU(0) storage
and fusing the sweep residual loop reduces mean elapsed time from $36.69$
to $35.84$ seconds and lowers peak resident memory by $35\%$, from
$0.195$ to $0.127$ GiB, while the final probability vectors agree to
$4.76\times10^{-16}$ in the $\ell_1$ norm.

Table~\ref{tab:reflection-cost} compares the baseline unreduced solver
with the reduced solver under reflection symmetry on both the
rare-switching calculation at cap $56$ and an adaptive FSP Schl\"ogl
integration. The adaptive benchmark integrates the Schl\"ogl model on
four voxels from initial state $(2,2,2,2)$ to $T=20$ using backward
Euler with $\Delta t=2$, cumulative FSP error budget
$\varepsilon_{\mathrm{FSP}}=10^{-5}$, and linear solve tolerance
$\varepsilon=10^{-10}$ under the acceptance rule of
Section~\ref{app:adaptive}. Both implementations dynamically expand the
population cap from ${N_{\max}}=8$ to ${N_{\max}}=82$ through identical
step rejections and execute $666$ total W-cycles.

\begin{table}[!htbp]
\centering
\caption{Comparison with exact reflection reduction on identical FSP problems
on four voxels. Stored unknowns report the dimension on the finest grid,
and elapsed time and peak resident memory are means of two alternating
runs.}
\label{tab:reflection-cost}
\begingroup\small\setlength{\tabcolsep}{4pt}
\begin{tabular}{lrr rr rr}
\toprule
& \multicolumn{2}{c}{Stored unknowns} & \multicolumn{2}{c}{Time (s)}
& \multicolumn{2}{c}{Memory (GiB)}\\
Problem & baseline & reduced & baseline & reduced & baseline & reduced\\
\midrule
Rare switching & 487,635 & 244,035 & 36.71 & 18.37 & 0.199 & 0.084\\
Adaptive Schl\"ogl & 2,123,555 & 1,062,229 & 60.03 & 30.50 & 0.903 & 0.486\\
\bottomrule
\end{tabular}\endgroup
\end{table}

Halving the number of stored unknowns through reflection symmetry reduces
mean elapsed time by factors of $2.00$ and $1.97$ (and CPU time from
$36.45$ to $17.92$ seconds and from $59.55$ to $29.94$ seconds), while
cutting peak memory by about half.\footnote{Two internal step timers in
one run were inconsistent with the external process duration and are
excluded from the summary, with all raw logs preserved in the data
archive.} The reconstructed fine probability vectors agree with the
unreduced solutions to $9.06\times10^{-15}$ and $3.95\times10^{-14}$ in
the $\ell_1$ norm, with all entries nonnegative.

Within the reduced solver on the rare-switching problem, omitting the
intermediate level of two voxels increases the total cycle count from
$1{,}215$ ($4\to2\to1$) to $2{,}179$ ($4\to1$) and raises elapsed time
from $18.37$ to $32.18$ seconds (a factor of $1.75$), while the final
probability vectors agree to $1.29\times10^{-15}$ in $\ell_1$. Thus
spatial coarsening and reflection reduction provide separate speedups
whenever the problem is invariant under reflection.

\FloatBarrier
\section{Spatial Front Propagation}\label{sec:spatial-front}
Because the FSP benchmarks on four voxels summarize spatial distributions
primarily through the global population marginal $k=\sum_v X_v$, we
include a separate trajectory simulation on a chain of $64$ voxels to
illustrate spatially inhomogeneous wave propagation in the bistable
Schl\"ogl model. This example is computed by Gillespie's exact
SSA~\cite{gillespie1977} and illustrates the spatial RDME dynamics
rather than the FSP linear solver.

Consider a chain of $64$ reflecting voxels of volume $\omega_v=20$ and
unit spacing, with diffusive jump rate $1$ per molecule to each adjacent
voxel. The local Schl\"ogl reaction constants are chosen as
$(c_1,c_2,c_3,c_4)=(0.17,0.03,0.5,0.4)$, corresponding to a
deterministic reaction drift $-0.005(u-2)(u-5)(u-10)$ for concentration
$u=X_v/\omega_v$ with stable equilibria at $u=2$ ($X_v=40$) and $u=10$
($X_v=200$) separated by an unstable barrier at $u=5$ ($X_v=100$). We
initialize the chain with a step interface placing $200$ molecules in
each of voxels $1,\ldots,16$ and $40$ molecules in each of voxels
$17,\ldots,64$, and simulate $256$ independent SSA trajectories to
$T=300$, recording voxel counts every $2$ time units.

\begin{figure}[!htbp]
\centering
\includegraphics[width=\linewidth]{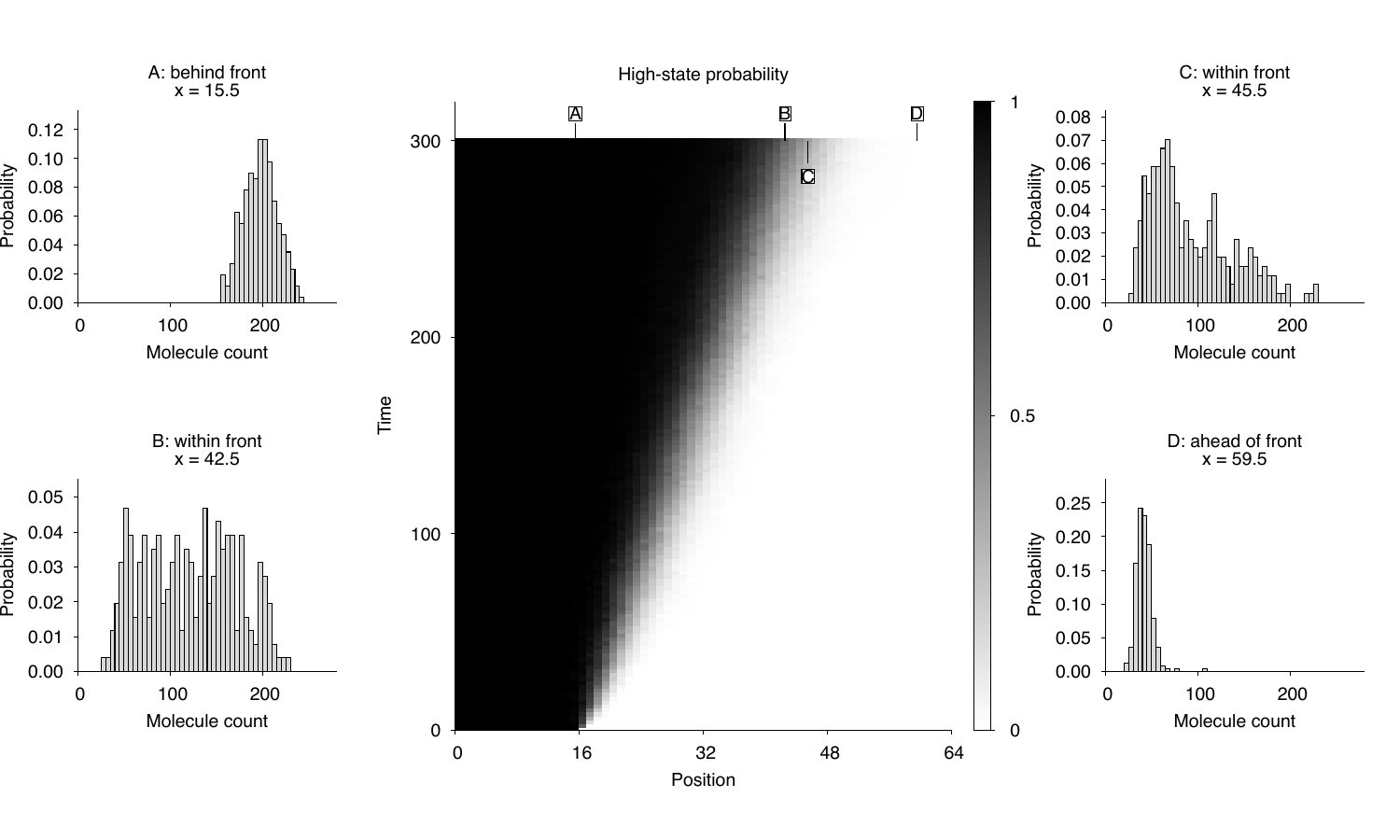}
\caption{Propagation of a stochastic Schl\"ogl front across a chain of 64
reflecting voxels estimated from 256 independent Gillespie SSA
trajectories. The central space--time map shows the empirical probability
$\mathbb P(X_v/20\ge5,t)$. Surrounding histograms A--D show the
normalized count distributions in individual voxels at $t=300$ (bin
width 5 molecules) behind the front (A), inside the interface region
(B and C), and ahead of the front (D).}
\label{fig:spatial-front}
\end{figure}

Figure~\ref{fig:spatial-front} plots the empirical probability
$\mathbb P(X_v/20\ge5,t)$ that a voxel has concentration at least $5$ as
the stochastic front advances from voxel $16$. At $T=300$, we also
extract the marginal count distributions at four representative voxel
locations A--D across the $256$ trajectories. Well behind the front
(voxel A) and well ahead of the front (voxel D), the local count
distributions are unimodal around $200$ and $40$ molecules, respectively.
Inside the transition zone (voxels B and C, where $\mathbb P(X_v/20\ge5)$
equals $0.625$ and $0.371$), the local voxel marginals broaden across
intermediate counts due to diffusive mixing and interface fluctuations,
rather than splitting into two sharply separated peaks within a single
voxel.

\FloatBarrier
\section*{Code Availability}
The solver and reproduction scripts are available at
\url{https://github.com/AdityaDendukuri/markovkit/tree/main/rdme}.

\Needspace{8\baselineskip}